\documentclass[aps, prl,twocolumn,superscriptaddress,showpacs,amsmath,amssymb, notitlepage]{revtex4-2}
\usepackage{amsfonts,hyperref}
\usepackage{float}
\usepackage{graphicx}
\usepackage{amsmath}
\usepackage{amssymb}

\usepackage{bm}

\begin{document}
\title{Quantum spin liquid phase in the Shastry-Sutherland model revealed by high-precision infinite projected entangled-pair states }
\author{Philippe \surname{Corboz}} \affiliation{Institute for Theoretical Physics, University of Amsterdam, Science Park 904, 1098 XH Amsterdam, The Netherlands}

\author{Yining \surname{Zhang}} \affiliation{Institute for Theoretical Physics, University of Amsterdam, Science Park 904, 1098 XH Amsterdam, The Netherlands}

\author{Boris \surname{Ponsioen}} \affiliation{Institute for Theoretical Physics, University of Amsterdam, Science Park 904, 1098 XH Amsterdam, The Netherlands}

\author{Fr\'ed\'eric \surname{Mila}} \affiliation{Institute of Physics, \'Ecole Polytechnique F\'ed\'erale de Lausanne (EPFL), CH-1015 Lausanne, Switzerland}

\begin{abstract}
The Shastry-Sutherland model is an effective model of the layered material SrCu$_2$(BO$_3$)$_2$, which exhibits an extremely rich phase diagram as a function of pressure and magnetic field. Motivated by the recent controversy regarding its phase diagram at zero magnetic field, we perform large-scale simulations based on infinite projected entangled-pair states (iPEPS), a two-dimensional tensor network ansatz  to represent the ground state directly in the thermodynamic limit. By employing  the latest optimization techniques, we obtain variational states with  lower energy than previous results obtained from other methods. Using systematic extrapolations to the exact infinite bond dimension limit, our simulations reveal a narrow quantum spin liquid phase  between the plaquette and antiferromagnetic phases in the range $0.785(5) \le J'/J \le 0.82(1)$.
\end{abstract}

\maketitle

\emph{Introduction.--} The frustrated  compound SrCu$_2$(BO$_3$)$_2$ is arguably one of the most fascinating quantum materials. It exhibits extremely rich physics when subjected to pressure and a magnetic field, including an intriguing sequence of magnetization plateaus~\cite{Kageyama99,Onizuka00,Kodama02,Takigawa04,levy08,Sebastian08,Jaime12,takigawa13,matsuda13,shi22, nomura23}, spin supersolid phases~\cite{matsuda13,shi22}, topological excitations~\cite{romhanyi15,mcclarty17}, a critical point at finite temperature~\cite{jimenez21}, and a proximate deconfined quantum critical point (DQCP)~\cite{cui23}. 
As a function of pressure, experiments have identified a phase transition between a dimer phase and a plaquette phase~\cite{waki07,haravifard16,Zayed17,sakurai18,bettler20,guo20,jimenez21,guo25}, followed by an antiferromagnetic (AF) phase at higher pressure~\cite{Zayed17,guo20,jimenez21,cui23,guo25}. While the possibility of a narrow quantum spin liquid or a DQCP in between the two latter phases has been suggested~\cite{cui23}, the latest specific heat data indicate a first-order phase transition~\cite{guo25}.  

Many of the properties of the compound are remarkably well captured by the Shastry-Sutherland model (SSM)~\cite{Shastry81,Kageyama99,Miyahara99,Miyahara03,corboz14_shastry,wessel18,wietek19,czarnik21,wang23,wang26,nyckees25}, a frustrated  $S=1/2$ spin model on an orthogonal dimer lattice, given by the Hamiltonian
\begin{equation}
H=J\sum_{\langle i,j \rangle}\bm S_{i}\cdot \bm S_{j}+J'\sum_{\langle \langle i,j \rangle\rangle}\bm S_{i}\cdot \bm S_{j}
\end{equation}
with $J$ and $J'$ the intra- and interdimer coupling strength, respectively. For small values of $J'/J$ the ground state is an exact product of dimer singlets~\cite{Shastry81}. In the $J'/J \rightarrow \infty $ limit, the model reduces to the square lattice Heisenberg model with an AF ground state. For intermediate $J'/J$ a valence-bond solid phase is realized, with strong bonds forming around plaquettes~\cite{Koga00,Takushima01,Laeuchli02,Corboz13_shastry}. 
An early infinite projected entangled-pair state (iPEPS) study~\cite{Corboz13_shastry} identified the range of the plaquette phase to be $0.675(2) < J'/J < 0.765(15)$, with transitions being strongly and weakly first order, respectively.

More recently, there have been new predictions regarding the location and nature of the second transition. A narrow quantum spin liquid phase  between plaquette and AF phase was predicted using the density matrix renormalization group (DMRG) on cylinders~\cite{yang22}, exact diagonalization (ED) and DMRG on periodic clusters~\cite{wang22},  a pseudofermion functional renormalization group (PFRG) approach~\cite{keles22},   neural-network quantum states (NQS)~\cite{viteritti25}, and variational Monte Carlo (VMC)~\cite{maity25}. Other DMRG~\cite{qian24} and iPEPS~\cite{xi23} studies found a weak first-order transition at a slightly higher value of $J'/J$. Finally, a DQCP was predicted in Ref.~\cite{lee19} based on field theory and infinite DMRG results, and in Ref.~\cite{liu24} based on variational Monte Carlo (VMC)  combined with finite PEPS (VMC-PEPS). Thus, a consensus on the phase diagram is still lacking, highlighting the need for additional high-precision data.

\begin{figure}[tb]
  \centering
  \includegraphics[width=1\linewidth]{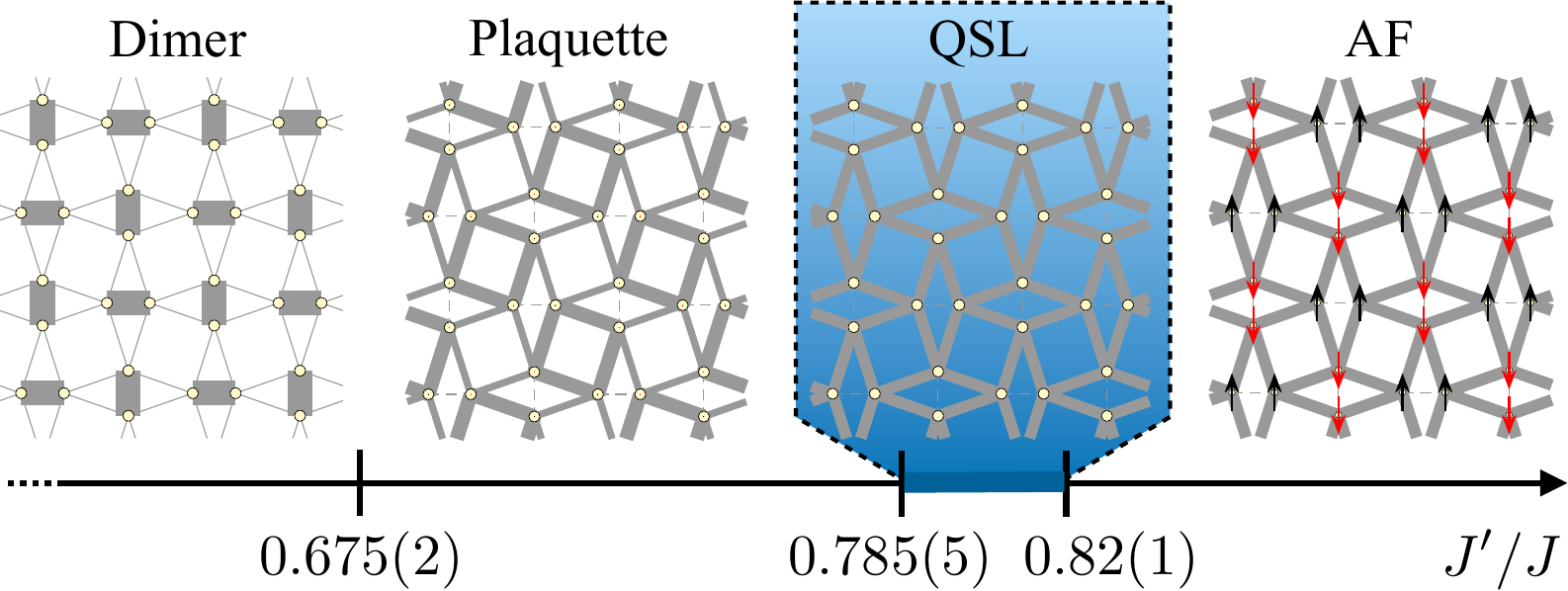}
  \caption{Phase diagram of the Shastry-Sutherland model obtained with iPEPS. A narrow quantum spin liquid (QSL) phase is found  between the plaquette and antiferromagnetic phases. The thickness of a bond is proportional to the negative  bond energy, and dashed lines indicate bonds with positive energies.
 }
  \label{fig:pd}
\end{figure}

In this Letter, we revisit the SSM using large-scale iPEPS simulations to resolve the controversial phase diagram. By leveraging the latest methodological advances, including more accurate and efficient optimization and systematic extrapolation techniques, we can simulate the model with unprecedented accuracy directly in the thermodynamic limit. Our results uncover a quantum spin liquid (QSL) phase within a narrow window $0.785(5) < J'/J < 0.82(1)$  between the plaquette and AF phases, as summarized in Fig.~\ref{fig:pd}. Our findings are consistent with the predictions from DMRG~\cite{yang22}, ED~\cite{wang22} and NQS~\cite{viteritti25} based on finite-size data. Besides this, we  demonstrate that iPEPS achieves, in the thermodynamic limit, lower energies than existing finite-size NQS calculations while using over an order of magnitude fewer variational parameters, demonstrating the relevance of the variational space defined by tensor networks and further establishing iPEPS as a state-of-the-art tool for studying frustrated spin systems, complementing  finite-size 2D approaches.

\emph{Method.--} 
An iPEPS is a two-dimensional tensor network ansatz to represent 2D ground states directly in the thermodynamic limit~\cite{verstraete2004,nishio2004,jordan2008}, which has been successfully applied to a broad range of strongly correlated systems, see e.g. Refs.~\cite{corboz14_tJ, nataf16, liao17, niesen17, chen18, lee18, jahromi18, niesen18, yamaguchi18, kshetrimayum19b, chung19, ponsioen19, lee20, gauthe20, hasik21, liu22b, peschke22, hasik22, ponsioen23b, weerda24, xu23b, hasik24,schmoll24}.
It consists of a unit cell of rank-5 tensors repeated on a square lattice, where each tensor has four auxiliary legs with bond dimension $D$ connecting the neighboring tensors and one physical leg with dimension $d$, representing the local Hilbert space. The accuracy of the ansatz can be systematically controlled by $D$. We consider two different setups: the dimer setup, which uses one tensor per dimer ($d=4$) and the single-site setup, which uses one tensor per site ($d=2$). To improve the efficiency, we exploit the U(1) spin symmetry of the model~\cite{singh2010,bauer2011}. The contraction of the tensor network is performed using the corner transfer matrix method~\cite{nishino1996, Orus2009, Corboz2011, corboz14_tJ}. Its accuracy is controlled by the environment bond dimension $\chi$, which we set sufficiently large (up to $\chi=500$ for $D=10$) to ensure that the contraction error of local observables is negligible. The tensors are optimized by minimizing the energy of the ansatz using automatic differentiation (AD)~\cite{liao19,ponsioen22}, which produces more accurate tensors than  methods based on imaginary time evolution~\cite{jiang2008,phien15,corboz16b}. 
To improve  performance, we use the recent extension of AD to the truncated singular value decomposition~\cite{francuz25}, enabling us to reach larger bond dimensions with AD than in previous iPEPS calculations~\cite{xi23}. The extrapolation of the energy to the infinite $D$ limit is based on the energy variance, which is computed using a newly developed technique based on a large-cell contraction~\cite{cortes25,arias24} (see also Ref.~\cite{vanderstraeten16}). In the End Matter, we provide more details on the methods.

\begin{figure}[tb]
  \centering
  \includegraphics[width=1\linewidth]{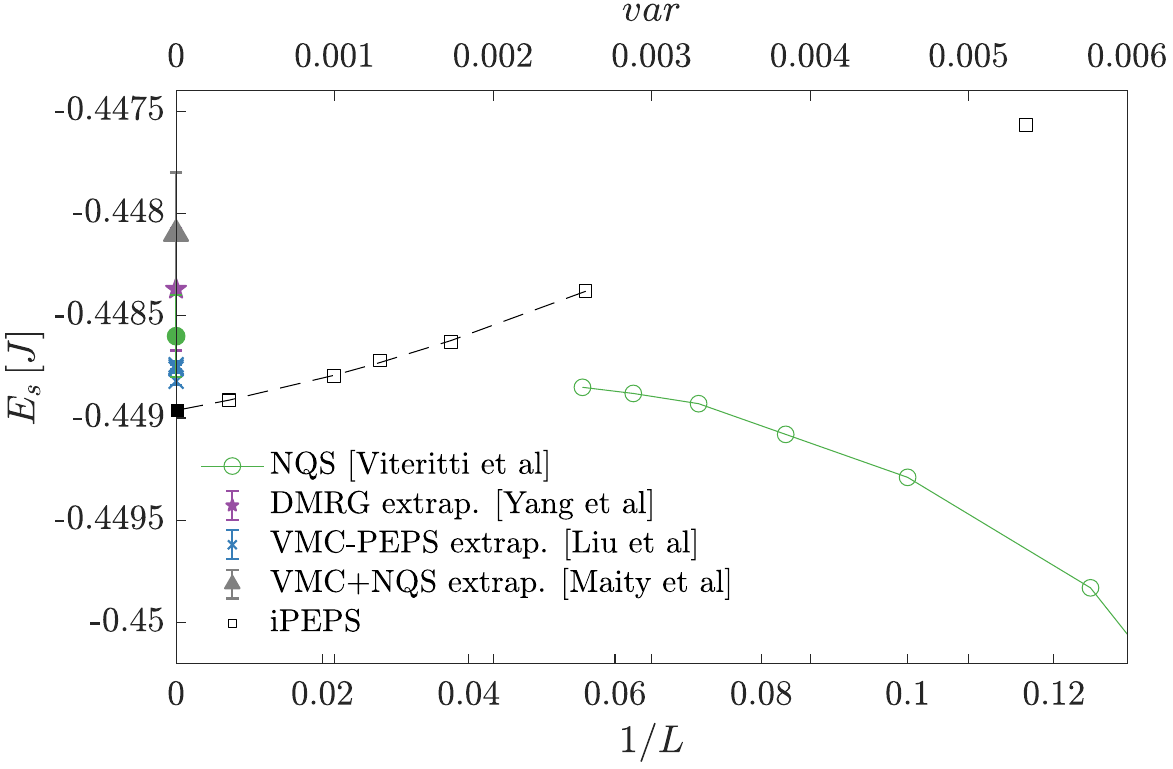}
  \caption{
Comparison of the energy per site $E_s$ obtained with different methods. The iPEPS data is plotted as a function of the variance of the energy per site. The $D=10$ iPEPS has the lowest variational energy and lies  close to the extrapolated value in the infinite $D$ limit.}
  \label{fig:bm}
\end{figure}

\emph{Benchmark comparison.--} 
To demonstrate the competitiveness of our approach, we first present a benchmark comparison of the energy per site, $E_s$, at $J'/J=0.8$ between our iPEPS results and data from NQS~\cite{viteritti25}, DMRG~\cite{yang22},  VMC-PEPS~\cite{liu24}, and VMC combined with NQS~\cite{maity25}. In Fig.~\ref{fig:bm}, the iPEPS energies in the thermodynamic limit are plotted as a function of variance of the energy per site, $var = (\langle H^2 \rangle - \langle H\rangle^2)/N$~\footnote{Extrapolations based on the bond dimension are generally not very accurate, because the functional behavior of E with respect to D is typically not smooth, and the functional form is unknown. In contrast, the energy as a function of the variance is much smoother, and close to convergence, the energy depends linearly on the variance. Here we have used a second-order polynomial fit to include corrections to the linear scaling}.
The lowest variational energy for $D=10$ with approximately 16'000 variational parameters is $-0.44891$  (in units of $J$), and the extrapolated value is  $-0.44896(2)$. 
The NQS data is based on a Vision Transformer (ViT) variational wave function with approximately 300'000 variational parameters on periodic $L\times L$ lattices. The energy  increases with system size, reaching $E_s=-0.44885 $ for the largest  size, with an extrapolated value of $-0.4486(2)$, which are both higher than the $D=10$ iPEPS result.

The DMRG data was obtained for $2L \times L$ cylinders with sizes up to $L=14$~\cite{yang22}, with energies above $-0.447$ (not visible in the plot) and an extrapolated value of $-0.44837(30)$. The VMC results are based on finite PEPS wavefunctions with open boundary conditions and a fixed bond dimension of $D=8$, for sizes up to $L=20$. The various extrapolations shown in Fig.~\ref{fig:bm} were obtained by evaluating the energy on different clusters. They lie within the range $[-0.44882,-0.44874]$, which is compatible with our $D=8$ iPEPS result but higher than the value in the infinite-$D$ limit. This indicates that,  to obtain accurate thermodynamic estimates based on finite PEPS, extrapolations in both $D$ and $L$ are required, rather than extrapolating only in $L$ at a fixed~$D$.

\begin{figure}[t!]
  \centering
  \includegraphics[width=1\linewidth]{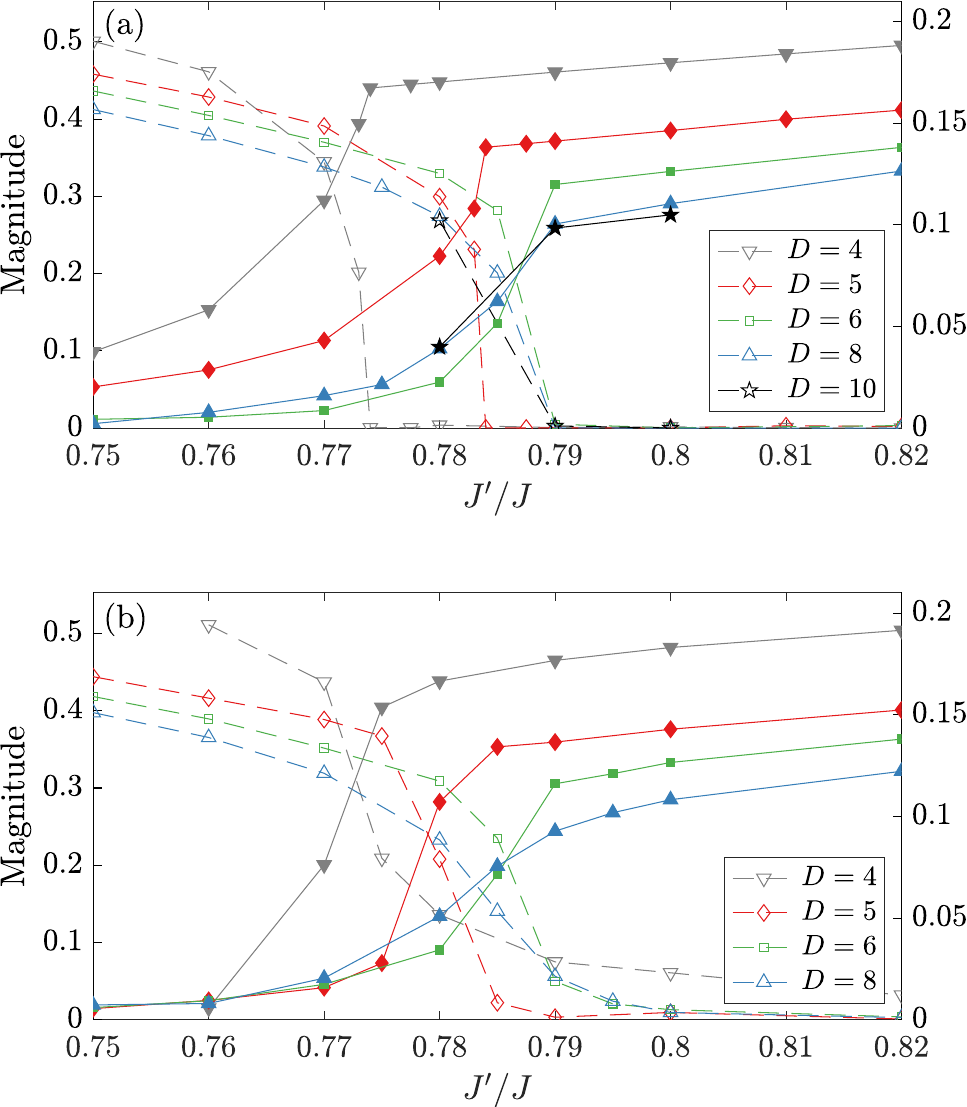}
  \caption{Plaquette order parameter $\Delta$ (open symbols, left axis) and local magnetic moment $m$ (filled symbols, right axis) for different bond dimensions as a function of $J'/J$, obtained with the dimer setup (a) and single-site setup (b). The  transition initially shifts to larger values of $J'/J$ with increasing $D$, but stabilizes in between $J'/J=[0.78,0.79]$ for $D\ge 6$.
 }
  \label{fig:ops}
\end{figure}

\emph{Phase transition at finite $D$.--} We next discuss the results for the transition between the plaquette phase and its adjacent phase at larger $J'/J$ for finite $D$, shown in Fig~\ref{fig:ops}(a) and Fig~\ref{fig:ops}(b) for the dimer and single-site setups, respectively. At finite $D$, the adjacent phase exhibits finite AF order, however, we will show further below that the order vanishes in the infinite $D$ limit, i.e. that the adjacent phase is in fact a QSL.
We consider the plaquette order parameter given by the difference between the high and low energy bonds, $\Delta  = \max(E_b) - \min(E_b)$ and the local magnetic moment, $m = \sqrt{\langle S^x \rangle^2+\langle S^y \rangle^2+\langle S^z \rangle^2}$.

In the dimer setup, a sudden drop in the plaquette order parameter is observed at the transition, consistent with a first order transition. The AF order parameter exhibits smoother behavior, suggesting either a continuous or weakly first-order transition. In the single-site setup, both order parameters show smooth behavior across the transition. In both setups, the location of the transition initially shifts from  $J'/J\sim0.77$ for $D=4$ to $J'/J\sim0.785$ for $D=6$ where it remains for larger values of $D$. 

\emph{Extrapolations based on constrained ans\"atze.--} The fact that at finite $D$  both order parameters are finite near the phase transition makes it challenging to accurately estimate  the energy of the pure individual phases. To resolve this problem, we consider two different constrained iPEPS ans\"atze based on the dimer setup, which respect the lattice and translational symmetries of the plaquette phase and the AF phase, respectively. The plaquette ansatz consists of a $2\times 2$ unit cell built from a single tensor $A$, which is rotated by 90 degrees between neighboring sites of the $2\times2$ cell. This ansatz is compatible with the plaquette order but not with the AF order; i.e., the AF order parameter is zero by construction. The AF ansatz is parametrized by a single tensor that is mirror symmetric with respect to the x- and y- axes on one sublattice, and by its SU(2) conjugate (with flipped spin), rotated by 90 degrees, on the other sublattice. This ansatz  can produce finite AF order but no plaquette order. We note that both ans\"atze are also compatible with a potential QSL phase.

\begin{figure}[tb!]
  \centering
  \hspace{0.3cm}
  \includegraphics[width=0.954\linewidth]{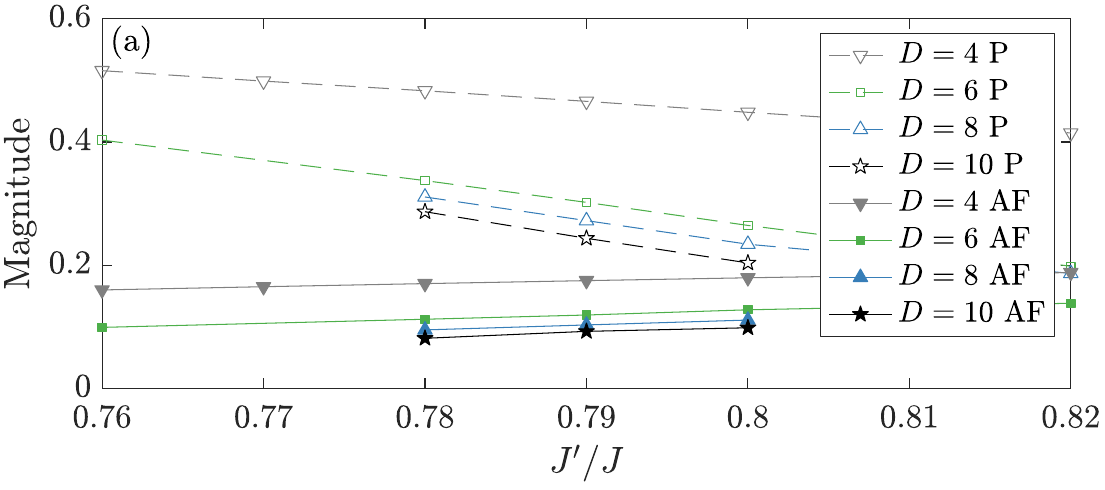}
    \vspace{0.15cm}
    \includegraphics[width=\linewidth]{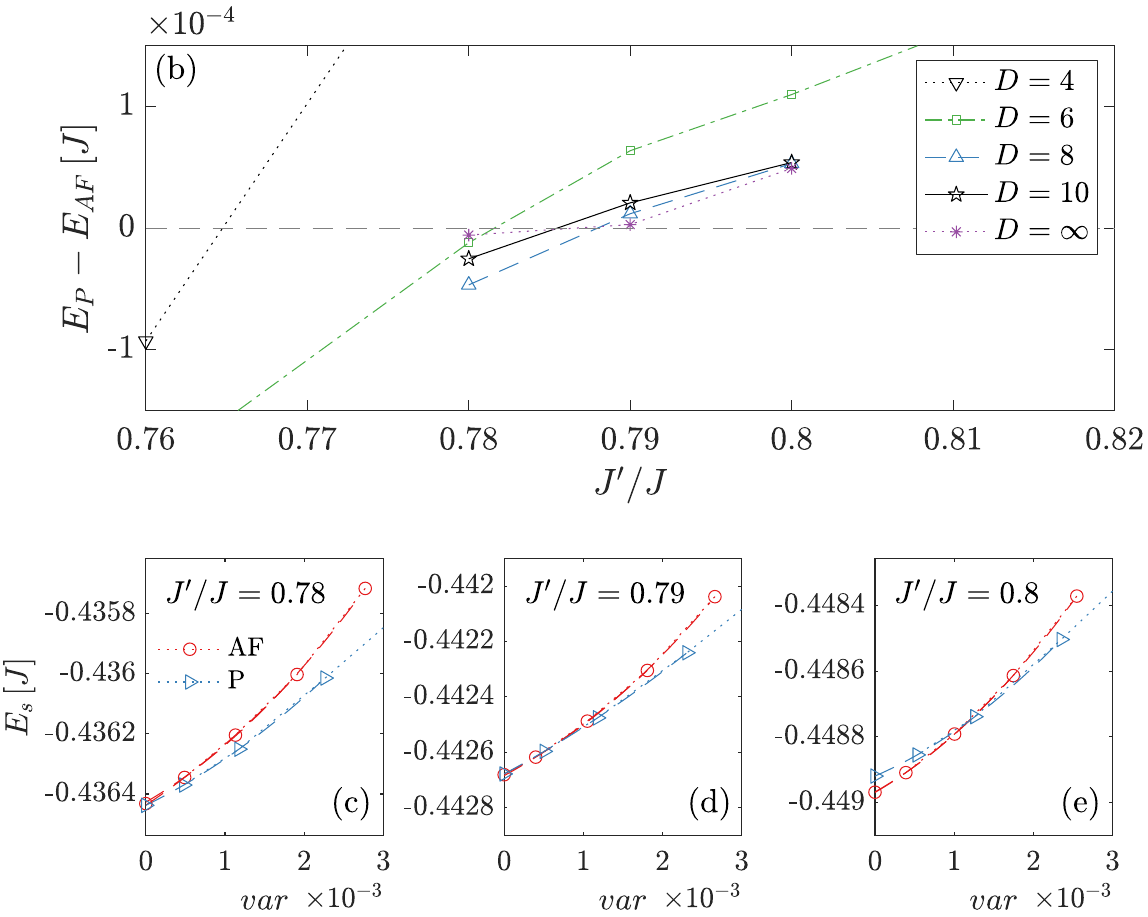}
  \caption{Results based on two constrained iPEPS ans\"atze: the plaquette (P) ansatz, which, by construction, can represent the plaquette order but not the AF order, and the AF ansatz, which can  represent the AF order but not the plaquette order. (a) Plaquette order parameter (open symbols) and local magnetic moment (filled symbols) as a function of $J'/J$ for different bond dimensions. (b) Energy difference between the plaquette state and the AF state. The location of the phase transition can be determined from the intersection with the dashed $E_P-E_{AF}=0$ line, which at large and infinite $D$, is located around $J'/J\sim0.785(5)$. (c)-(e) Energies of the two competing states for three different values of $J'/J$ around the transition as a function of the energy variance per site.}
  \label{fig:constrained}
\end{figure}

\begin{figure}[tbh]
  \centering
    \includegraphics[width=1\linewidth]{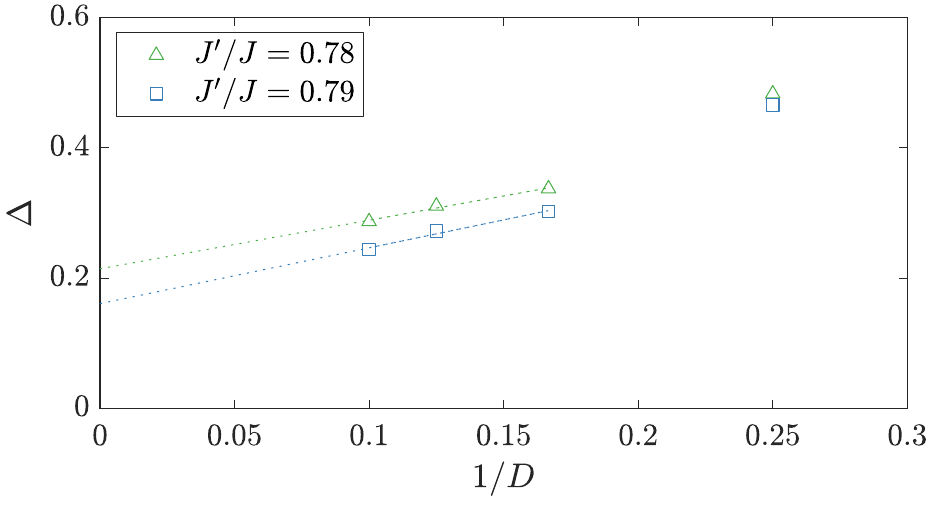}
  \caption{Plaquette order parameter $\Delta$ obtained with the plaquette ansatz as a function of $1/D$, showing that the order parameter does not vanish at the transition, i.e., it exhibits a discontinuous drop from a finite value in the plaquette phase to zero in the QSL phase, indicating a first-order phase transition. }
  \label{fig:pop}
\end{figure}

By using these constrained ans\"atze, we can accurately determine the energies of the individual phases and locate the transition from the intersection of their energies. Figure~\ref{fig:constrained}(a) shows the order parameters in the vicinity of the transition for the two ans\"atze at different $D$, illustrating that we can obtain clean representations of the two phases on both sides of the phase transition. 
Figure~\ref{fig:constrained}(b) shows the energy difference between the plaquette and AF states, from which the location of the phase transition can be determined. As previously observed, the phase transition shifts to larger values with increasing $D$ but stabilizes around $J'/J\sim0.785$ at large $D$, also when extrapolating the energies to the infinite $D$ limit based on the energy variance~\footnote{The extrapolations of the energies based on a second order polynomial, using the three largest $D$ values for the plaquette state. For the AF state we took the average of the extrapolated values based on the three and four largest $D$ values}, see Figs.~\ref{fig:constrained}(c)-(e). The extrapolated energies are found to be extremely close in the range $J'/J \in [0.78,0.79]$, with an energy difference below $10^{-5} J$. 
Finally, we find that the plaquette order parameter remains finite around the transition in the infinite $D$ limit (see Fig.~\ref{fig:pop} in the End Matter), which, along with the small angle between the intersecting energies at the transition, suggests that the transition is weakly first order~\footnote{A weak first-order transition refers to a discontinuous transition that is close to being continuous. It typically exhibits only mild hysteresis around the transition point; the discontinuity in the order parameter is small, and the energy curves of the two states intersect at a shallow angle~\cite{demidio23}}.

Thus, based on these results, we conclude that the plaquette phase persists up to a value $J'/J= 0.785(5)$, which is compatible with the prediction of DMRG from Ref.~\cite{yang22} ($J'/J=0.788(2)$) and Ref.~\cite{qian24} ($J'/J=0.785(5)$), ED~\cite{wang22} ($J'/J=0.789(4)$), and NQS ($J'/J\sim0.78$), but incompatible with  VMC-PEPS ($J'/J= 0.828(5)$), which may be due to the fact that those results were based on finite-$D$ data only with open boundary conditions. In the End Matter, we investigate the effect of the open boundary conditions based on a modified SSM model, and we find that they naturally lead to a shift of the phase boundary to larger $J'/J$ values.

\emph{Finite correlation length scaling.--}
We next turn our focus on the AF order parameter $m$ and perform a systematic finite correlation length scaling (FCLS) analysis~\cite{corboz18,rader18,hasik21,vanhecke22} to determine its value in the infinite $D$ limit. The main idea of FCLS~\cite{tagliacozzo08,pollmann2009} for gapless systems is similar to conventional finite-size scaling, except that the system size is replaced by the dominant correlation length $\xi_D$ extracted from the iPEPS ansatz for each value of $D$. For the square lattice Heisenberg model, which can be adiabatically connected to the AF phase in the SSM, the asymptotic behavior of the order parameter is $m^2(\xi_D) = m_{\infty} +\alpha/\xi_D + \mathcal{O}(1/\xi_D^2)$~\cite{hasenfratz93}. To estimate $\xi_D$ we employ the extrapolation technique from Ref.~\cite{rams18}.

\begin{figure}[tb]
  \centering
  \includegraphics[width=1\linewidth]{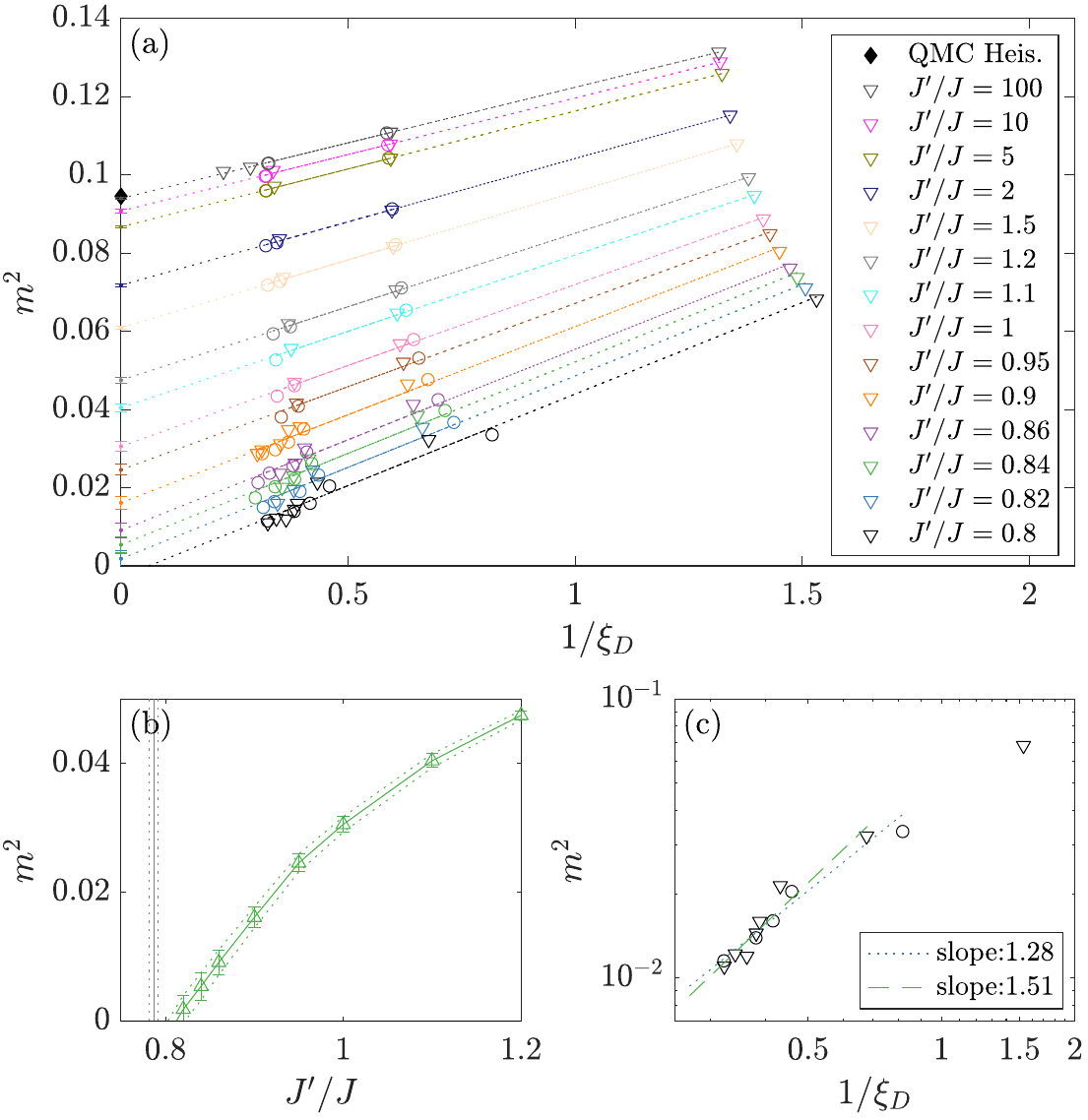}
  \caption{Finite correlation length scaling analysis of the squared  local magnetic moment. (a) Linear extrapolations of the $m^2$ data as a function of the inverse effective correlation length $\xi_D$ for different values of $J'/J$, based on the combined data from the dimer setup (triangles) and the single-site setup (circles). The QMC result~\cite{Sandvik2010} for the square-lattice Heisenberg model in the infinite $J'/J$ limit is indicated by the black diamond. (b) Extrapolated value of $m^2$, which vanishes for $J'/J\lesssim 0.812$, i.e. before the onset of the plaquette phase, indicated by the vertical grey line. 
  (c) Log-log plot of the $m^2$ data in the QSL phase for $J'/J=0.8,$ with linear fits based on two different data ranges, based on the combined data from the dimer setup (triangles) and the single-site setup (circles).
  }
  \label{fig:fcls}
\end{figure}

Figure.~\ref{fig:fcls}(a) shows the data for $m^2$ obtained with both the dimer and single-site setup as a function of inverse $\xi_D$. In the large $J'/J=100$ limit, the extrapolated result is compatible with the value from the square lattice Heisenberg model~\cite{Sandvik10}, as expected. Upon decreasing $J'/J$ the local magnetic moment decreases and vanishes around $J'/J\sim 0.812$, before reaching the onset of the plaquette phase. Omitting the smallest bond dimension value $D=3$ in the linear extrapolation yields a slightly higher value of $J'/J\sim 0.827$.

\emph{Discussion.--}
The vanishing local magnetic moment $m$ and the absence of lattice and translational symmetry breaking above the plaquette phase up to $J'/J = 0.82(1)$  are compatible with a QSL phase, consistent with previous results from DMRG~\cite{yang22} ($J'/J =0.820(2)$), ED~\cite{wang22} ($J'/J =0.826(3)$), PFRG~\cite{keles22} ($J'/J\sim0.82$), NQS~\cite{viteritti25} ($J'/J \sim0.82$),  and VMC~\cite{maity25}, which predicted an algebraic QSL~\footnote{A different conclusion was obtained in another iPEPS study~\cite{xi23}, however, the results were limited to $D$ values up to 7, and extrapolations of $m$ were performed based on a $1/D$ extrapolation instead of a more accurate FCLS analysis.}.
The fact that the order parameter remains finite at finite $D$ and only vanishes in the infinite $D$ limit is compatible with an algebraic quantum spin liquid rather than a gapped one. A similar behavior was observed in an iPEPS study of the $J_1$-$J_2$ Heisenberg model~\cite{hasik21} exhibiting a gapless QSL~\cite{hu13,richter15,wang18,hering19}. By fitting our data at $J'/J=0.8$ to the expected power-law scaling form $m^2(\xi_D) \sim \xi_D^{-(1+\eta)}$~\cite{nomura21,viteritti25}, we extract values for $\eta$ in the interval $[0.3, 0.5]$, depending on the fitting range,  compatible with the estimates $\eta=0.3$ from DMRG~\cite{yang22} and NQS~\cite{viteritti25}, and $\eta=0.5$ from VMC~\cite{maity25}.

In summary,  we revisited the phase diagram of the SSM based on  high-precision iPEPS simulations and identified a QSL phase in the range $ 0.785(5) \le J'/J  \le 0.82(1)$. 
Using constrained iPEPS ans\"atze enabled us to accurately determine the transition between the plaquette and QSL phase from the intersection of their  energies extrapolated as a function of the energy variance. From the discontinuous change in the plaquette order parameter at the transition, we concluded that the transition is first-order. 
By performing a systematic FCLS analysis of the AF order parameter, we located the continuous transition between the QSL and AF phases. Our data complement previous finite-size variational data and further support the algebraic nature of the QSL.

As a future direction, it will be interesting to investigate the robustness of our findings when incorporating  additional interactions beyond the standard SSM, such as a weak interlayer coupling~\cite{ueda99,koga00c,Miyahara03,vlaar23b,fogh24}, Dzyaloshinskii-Moriya interactions~\cite{Cepas01,Miyahara03,romhanyi11}, or small distortions in the coupling strengths~\cite{boos19}. In the End Matter, we present one example of a modified SSM~\cite{liu24} with a small additional interaction perpendicular to the dimer interaction, for which we find that the narrow QSL phase persists.

\acknowledgments
This project has received funding from the European Research Council (ERC) under the European Union's Horizon 2020 research and innovation programme (grant agreement No. 101001604).

\bibliographystyle{apsrev4-2}
\bibliography{../bib/refs.bib}

\begin{thebibliography}{113}%
\makeatletter
\providecommand \@ifxundefined [1]{%
 \@ifx{#1\undefined}
}%
\providecommand \@ifnum [1]{%
 \ifnum #1\expandafter \@firstoftwo
 \else \expandafter \@secondoftwo
 \fi
}%
\providecommand \@ifx [1]{%
 \ifx #1\expandafter \@firstoftwo
 \else \expandafter \@secondoftwo
 \fi
}%
\providecommand \natexlab [1]{#1}%
\providecommand \enquote  [1]{``#1''}%
\providecommand \bibnamefont  [1]{#1}%
\providecommand \bibfnamefont [1]{#1}%
\providecommand \citenamefont [1]{#1}%
\providecommand \href@noop [0]{\@secondoftwo}%
\providecommand \href [0]{\begingroup \@sanitize@url \@href}%
\providecommand \@href[1]{\@@startlink{#1}\@@href}%
\providecommand \@@href[1]{\endgroup#1\@@endlink}%
\providecommand \@sanitize@url [0]{\catcode `\\12\catcode `\$12\catcode
  `\&12\catcode `\#12\catcode `\^12\catcode `\_12\catcode `\%12\relax}%
\providecommand \@@startlink[1]{}%
\providecommand \@@endlink[0]{}%
\providecommand \url  [0]{\begingroup\@sanitize@url \@url }%
\providecommand \@url [1]{\endgroup\@href {#1}{\urlprefix }}%
\providecommand \urlprefix  [0]{URL }%
\providecommand \Eprint [0]{\href }%
\providecommand \doibase [0]{https://doi.org/}%
\providecommand \selectlanguage [0]{\@gobble}%
\providecommand \bibinfo  [0]{\@secondoftwo}%
\providecommand \bibfield  [0]{\@secondoftwo}%
\providecommand \translation [1]{[#1]}%
\providecommand \BibitemOpen [0]{}%
\providecommand \bibitemStop [0]{}%
\providecommand \bibitemNoStop [0]{.\EOS\space}%
\providecommand \EOS [0]{\spacefactor3000\relax}%
\providecommand \BibitemShut  [1]{\csname bibitem#1\endcsname}%
\let\auto@bib@innerbib\@empty
\bibitem [{\citenamefont {Kageyama}\ \emph {et~al.}(1999)\citenamefont
  {Kageyama}, \citenamefont {Yoshimura}, \citenamefont {Stern}, \citenamefont
  {Mushnikov}, \citenamefont {Onizuka}, \citenamefont {Kato}, \citenamefont
  {Kosuge}, \citenamefont {Slichter}, \citenamefont {Goto},\ and\ \citenamefont
  {Ueda}}]{Kageyama99}%
  \BibitemOpen
  \bibfield  {author} {\bibinfo {author} {\bibfnamefont {H.}~\bibnamefont
  {Kageyama}}, \bibinfo {author} {\bibfnamefont {K.}~\bibnamefont {Yoshimura}},
  \bibinfo {author} {\bibfnamefont {R.}~\bibnamefont {Stern}}, \bibinfo
  {author} {\bibfnamefont {N.~V.}\ \bibnamefont {Mushnikov}}, \bibinfo {author}
  {\bibfnamefont {K.}~\bibnamefont {Onizuka}}, \bibinfo {author} {\bibfnamefont
  {M.}~\bibnamefont {Kato}}, \bibinfo {author} {\bibfnamefont {K.}~\bibnamefont
  {Kosuge}}, \bibinfo {author} {\bibfnamefont {C.~P.}\ \bibnamefont
  {Slichter}}, \bibinfo {author} {\bibfnamefont {T.}~\bibnamefont {Goto}},\
  and\ \bibinfo {author} {\bibfnamefont {Y.}~\bibnamefont {Ueda}},\ }\href
  {https://doi.org/10.1103/PhysRevLett.82.3168} {\bibfield  {journal} {\bibinfo
   {journal} {Phys. Rev. Lett.}\ }\textbf {\bibinfo {volume} {82}},\ \bibinfo
  {pages} {3168} (\bibinfo {year} {1999})}\BibitemShut {NoStop}%
\bibitem [{\citenamefont {Onizuka}\ \emph {et~al.}(2000)\citenamefont
  {Onizuka}, \citenamefont {Kageyama}, \citenamefont {Narumi}, \citenamefont
  {Kindo}, \citenamefont {Ueda},\ and\ \citenamefont {Goto}}]{Onizuka00}%
  \BibitemOpen
  \bibfield  {author} {\bibinfo {author} {\bibfnamefont {K.}~\bibnamefont
  {Onizuka}}, \bibinfo {author} {\bibfnamefont {H.}~\bibnamefont {Kageyama}},
  \bibinfo {author} {\bibfnamefont {Y.}~\bibnamefont {Narumi}}, \bibinfo
  {author} {\bibfnamefont {K.}~\bibnamefont {Kindo}}, \bibinfo {author}
  {\bibfnamefont {Y.}~\bibnamefont {Ueda}},\ and\ \bibinfo {author}
  {\bibfnamefont {T.}~\bibnamefont {Goto}},\ }\href
  {https://doi.org/10.1143/JPSJ.69.1016} {\bibfield  {journal} {\bibinfo
  {journal} {J. Phys. Soc. Jpn.}\ }\textbf {\bibinfo {volume} {69}},\ \bibinfo
  {pages} {1016} (\bibinfo {year} {2000})}\BibitemShut {NoStop}%
\bibitem [{\citenamefont {Kodama}\ \emph {et~al.}(2002)\citenamefont {Kodama},
  \citenamefont {Takigawa}, \citenamefont {Horvati{\'c}}, \citenamefont
  {Berthier}, \citenamefont {Kageyama}, \citenamefont {Ueda}, \citenamefont
  {Miyahara}, \citenamefont {Becca},\ and\ \citenamefont {Mila}}]{Kodama02}%
  \BibitemOpen
  \bibfield  {author} {\bibinfo {author} {\bibfnamefont {K.}~\bibnamefont
  {Kodama}}, \bibinfo {author} {\bibfnamefont {M.}~\bibnamefont {Takigawa}},
  \bibinfo {author} {\bibfnamefont {M.}~\bibnamefont {Horvati{\'c}}}, \bibinfo
  {author} {\bibfnamefont {C.}~\bibnamefont {Berthier}}, \bibinfo {author}
  {\bibfnamefont {H.}~\bibnamefont {Kageyama}}, \bibinfo {author}
  {\bibfnamefont {Y.}~\bibnamefont {Ueda}}, \bibinfo {author} {\bibfnamefont
  {S.}~\bibnamefont {Miyahara}}, \bibinfo {author} {\bibfnamefont
  {F.}~\bibnamefont {Becca}},\ and\ \bibinfo {author} {\bibfnamefont
  {F.}~\bibnamefont {Mila}},\ }\href {https://doi.org/10.1126/science.1075045}
  {\bibfield  {journal} {\bibinfo  {journal} {Science}\ }\textbf {\bibinfo
  {volume} {298}},\ \bibinfo {pages} {395} (\bibinfo {year}
  {2002})}\BibitemShut {NoStop}%
\bibitem [{\citenamefont {Takigawa}\ \emph {et~al.}(2004)\citenamefont
  {Takigawa}, \citenamefont {Kodama}, \citenamefont {Horvati{\'c}},
  \citenamefont {Berthier}, \citenamefont {Kageyama}, \citenamefont {Ueda},
  \citenamefont {Miyahara}, \citenamefont {Becca},\ and\ \citenamefont
  {Mila}}]{Takigawa04}%
  \BibitemOpen
  \bibfield  {author} {\bibinfo {author} {\bibfnamefont {M.}~\bibnamefont
  {Takigawa}}, \bibinfo {author} {\bibfnamefont {K.}~\bibnamefont {Kodama}},
  \bibinfo {author} {\bibfnamefont {M.}~\bibnamefont {Horvati{\'c}}}, \bibinfo
  {author} {\bibfnamefont {C.}~\bibnamefont {Berthier}}, \bibinfo {author}
  {\bibfnamefont {H.}~\bibnamefont {Kageyama}}, \bibinfo {author}
  {\bibfnamefont {Y.}~\bibnamefont {Ueda}}, \bibinfo {author} {\bibfnamefont
  {S.}~\bibnamefont {Miyahara}}, \bibinfo {author} {\bibfnamefont
  {F.}~\bibnamefont {Becca}},\ and\ \bibinfo {author} {\bibfnamefont
  {F.}~\bibnamefont {Mila}},\ }\href
  {https://doi.org/10.1016/j.physb.2004.01.014} {\bibfield  {journal} {\bibinfo
   {journal} {Physica B: Condensed Matter}\ }\textbf {\bibinfo {volume}
  {346{\textendash}347}},\ \bibinfo {pages} {27} (\bibinfo {year}
  {2004})}\BibitemShut {NoStop}%
\bibitem [{\citenamefont {Levy}\ \emph {et~al.}(2008)\citenamefont {Levy},
  \citenamefont {Sheikin}, \citenamefont {Berthier}, \citenamefont
  {Horvati{\'c}}, \citenamefont {Takigawa}, \citenamefont {Kageyama},
  \citenamefont {Waki},\ and\ \citenamefont {Ueda}}]{levy08}%
  \BibitemOpen
  \bibfield  {author} {\bibinfo {author} {\bibfnamefont {F.}~\bibnamefont
  {Levy}}, \bibinfo {author} {\bibfnamefont {I.}~\bibnamefont {Sheikin}},
  \bibinfo {author} {\bibfnamefont {C.}~\bibnamefont {Berthier}}, \bibinfo
  {author} {\bibfnamefont {M.}~\bibnamefont {Horvati{\'c}}}, \bibinfo {author}
  {\bibfnamefont {M.}~\bibnamefont {Takigawa}}, \bibinfo {author}
  {\bibfnamefont {H.}~\bibnamefont {Kageyama}}, \bibinfo {author}
  {\bibfnamefont {T.}~\bibnamefont {Waki}},\ and\ \bibinfo {author}
  {\bibfnamefont {Y.}~\bibnamefont {Ueda}},\ }\href
  {https://doi.org/10.1209/0295-5075/81/67004} {\bibfield  {journal} {\bibinfo
  {journal} {{EPL} (Europhysics Letters)}\ }\textbf {\bibinfo {volume} {81}},\
  \bibinfo {pages} {67004} (\bibinfo {year} {2008})}\BibitemShut {NoStop}%
\bibitem [{\citenamefont {Sebastian}\ \emph {et~al.}(2008)\citenamefont
  {Sebastian}, \citenamefont {Harrison}, \citenamefont {Sengupta},
  \citenamefont {Batista}, \citenamefont {Francoual}, \citenamefont {Palm},
  \citenamefont {Murphy}, \citenamefont {Marcano}, \citenamefont {Dabkowska},\
  and\ \citenamefont {Gaulin}}]{Sebastian08}%
  \BibitemOpen
  \bibfield  {author} {\bibinfo {author} {\bibfnamefont {S.~E.}\ \bibnamefont
  {Sebastian}}, \bibinfo {author} {\bibfnamefont {N.}~\bibnamefont {Harrison}},
  \bibinfo {author} {\bibfnamefont {P.}~\bibnamefont {Sengupta}}, \bibinfo
  {author} {\bibfnamefont {C.~D.}\ \bibnamefont {Batista}}, \bibinfo {author}
  {\bibfnamefont {S.}~\bibnamefont {Francoual}}, \bibinfo {author}
  {\bibfnamefont {E.}~\bibnamefont {Palm}}, \bibinfo {author} {\bibfnamefont
  {T.}~\bibnamefont {Murphy}}, \bibinfo {author} {\bibfnamefont
  {N.}~\bibnamefont {Marcano}}, \bibinfo {author} {\bibfnamefont {H.~A.}\
  \bibnamefont {Dabkowska}},\ and\ \bibinfo {author} {\bibfnamefont {B.~D.}\
  \bibnamefont {Gaulin}},\ }\href {https://doi.org/10.1073/pnas.0804320105}
  {\bibfield  {journal} {\bibinfo  {journal} {PNAS}\ }\textbf {\bibinfo
  {volume} {105}},\ \bibinfo {pages} {20157} (\bibinfo {year}
  {2008})}\BibitemShut {NoStop}%
\bibitem [{\citenamefont {Jaime}\ \emph {et~al.}(2012)\citenamefont {Jaime},
  \citenamefont {Daou}, \citenamefont {Crooker}, \citenamefont {Weickert},
  \citenamefont {Uchida}, \citenamefont {Feiguin}, \citenamefont {Batista},
  \citenamefont {Dabkowska},\ and\ \citenamefont {Gaulin}}]{Jaime12}%
  \BibitemOpen
  \bibfield  {author} {\bibinfo {author} {\bibfnamefont {M.}~\bibnamefont
  {Jaime}}, \bibinfo {author} {\bibfnamefont {R.}~\bibnamefont {Daou}},
  \bibinfo {author} {\bibfnamefont {S.~A.}\ \bibnamefont {Crooker}}, \bibinfo
  {author} {\bibfnamefont {F.}~\bibnamefont {Weickert}}, \bibinfo {author}
  {\bibfnamefont {A.}~\bibnamefont {Uchida}}, \bibinfo {author} {\bibfnamefont
  {A.~E.}\ \bibnamefont {Feiguin}}, \bibinfo {author} {\bibfnamefont {C.~D.}\
  \bibnamefont {Batista}}, \bibinfo {author} {\bibfnamefont {H.~A.}\
  \bibnamefont {Dabkowska}},\ and\ \bibinfo {author} {\bibfnamefont {B.~D.}\
  \bibnamefont {Gaulin}},\ }\href {https://doi.org/10.1073/pnas.1200743109}
  {\bibfield  {journal} {\bibinfo  {journal} {PNAS}\ }\textbf {\bibinfo
  {volume} {109}},\ \bibinfo {pages} {12404} (\bibinfo {year}
  {2012})}\BibitemShut {NoStop}%
\bibitem [{\citenamefont {Takigawa}\ \emph {et~al.}(2013)\citenamefont
  {Takigawa}, \citenamefont {Horvati{\'c}}, \citenamefont {Waki}, \citenamefont
  {Kr{\"a}mer}, \citenamefont {Berthier}, \citenamefont {L{\'e}vy-Bertrand},
  \citenamefont {Sheikin}, \citenamefont {Kageyama}, \citenamefont {Ueda},\
  and\ \citenamefont {Mila}}]{takigawa13}%
  \BibitemOpen
  \bibfield  {author} {\bibinfo {author} {\bibfnamefont {M.}~\bibnamefont
  {Takigawa}}, \bibinfo {author} {\bibfnamefont {M.}~\bibnamefont
  {Horvati{\'c}}}, \bibinfo {author} {\bibfnamefont {T.}~\bibnamefont {Waki}},
  \bibinfo {author} {\bibfnamefont {S.}~\bibnamefont {Kr{\"a}mer}}, \bibinfo
  {author} {\bibfnamefont {C.}~\bibnamefont {Berthier}}, \bibinfo {author}
  {\bibfnamefont {F.}~\bibnamefont {L{\'e}vy-Bertrand}}, \bibinfo {author}
  {\bibfnamefont {I.}~\bibnamefont {Sheikin}}, \bibinfo {author} {\bibfnamefont
  {H.}~\bibnamefont {Kageyama}}, \bibinfo {author} {\bibfnamefont
  {Y.}~\bibnamefont {Ueda}},\ and\ \bibinfo {author} {\bibfnamefont
  {F.}~\bibnamefont {Mila}},\ }\href
  {https://doi.org/10.1103/PhysRevLett.110.067210} {\bibfield  {journal}
  {\bibinfo  {journal} {Phys. Rev. Lett.}\ }\textbf {\bibinfo {volume} {110}},\
  \bibinfo {pages} {067210} (\bibinfo {year} {2013})}\BibitemShut {NoStop}%
\bibitem [{\citenamefont {Matsuda}\ \emph {et~al.}(2013)\citenamefont
  {Matsuda}, \citenamefont {Abe}, \citenamefont {Takeyama}, \citenamefont
  {Kageyama}, \citenamefont {Corboz}, \citenamefont {Honecker}, \citenamefont
  {Manmana}, \citenamefont {Foltin}, \citenamefont {Schmidt},\ and\
  \citenamefont {Mila}}]{matsuda13}%
  \BibitemOpen
  \bibfield  {author} {\bibinfo {author} {\bibfnamefont {Y.~H.}\ \bibnamefont
  {Matsuda}}, \bibinfo {author} {\bibfnamefont {N.}~\bibnamefont {Abe}},
  \bibinfo {author} {\bibfnamefont {S.}~\bibnamefont {Takeyama}}, \bibinfo
  {author} {\bibfnamefont {H.}~\bibnamefont {Kageyama}}, \bibinfo {author}
  {\bibfnamefont {P.}~\bibnamefont {Corboz}}, \bibinfo {author} {\bibfnamefont
  {A.}~\bibnamefont {Honecker}}, \bibinfo {author} {\bibfnamefont {S.~R.}\
  \bibnamefont {Manmana}}, \bibinfo {author} {\bibfnamefont {G.~R.}\
  \bibnamefont {Foltin}}, \bibinfo {author} {\bibfnamefont {K.~P.}\
  \bibnamefont {Schmidt}},\ and\ \bibinfo {author} {\bibfnamefont
  {F.}~\bibnamefont {Mila}},\ }\href
  {https://doi.org/10.1103/PhysRevLett.111.137204} {\bibfield  {journal}
  {\bibinfo  {journal} {Phys. Rev. Lett.}\ }\textbf {\bibinfo {volume} {111}},\
  \bibinfo {pages} {137204} (\bibinfo {year} {2013})}\BibitemShut {NoStop}%
\bibitem [{\citenamefont {Shi}\ \emph {et~al.}(2022)\citenamefont {Shi},
  \citenamefont {Dissanayake}, \citenamefont {Corboz}, \citenamefont
  {Steinhardt}, \citenamefont {Graf}, \citenamefont {Silevitch}, \citenamefont
  {Dabkowska}, \citenamefont {Rosenbaum}, \citenamefont {Mila},\ and\
  \citenamefont {Haravifard}}]{shi22}%
  \BibitemOpen
  \bibfield  {author} {\bibinfo {author} {\bibfnamefont {Z.}~\bibnamefont
  {Shi}}, \bibinfo {author} {\bibfnamefont {S.}~\bibnamefont {Dissanayake}},
  \bibinfo {author} {\bibfnamefont {P.}~\bibnamefont {Corboz}}, \bibinfo
  {author} {\bibfnamefont {W.}~\bibnamefont {Steinhardt}}, \bibinfo {author}
  {\bibfnamefont {D.}~\bibnamefont {Graf}}, \bibinfo {author} {\bibfnamefont
  {D.~M.}\ \bibnamefont {Silevitch}}, \bibinfo {author} {\bibfnamefont {H.~A.}\
  \bibnamefont {Dabkowska}}, \bibinfo {author} {\bibfnamefont {T.~F.}\
  \bibnamefont {Rosenbaum}}, \bibinfo {author} {\bibfnamefont {F.}~\bibnamefont
  {Mila}},\ and\ \bibinfo {author} {\bibfnamefont {S.}~\bibnamefont
  {Haravifard}},\ }\href {https://doi.org/10.1038/s41467-022-30036-w}
  {\bibfield  {journal} {\bibinfo  {journal} {Nat. Comm.}\ }\textbf {\bibinfo
  {volume} {13}},\ \bibinfo {pages} {2301} (\bibinfo {year}
  {2022})}\BibitemShut {NoStop}%
\bibitem [{\citenamefont {Nomura}\ \emph {et~al.}(2023)\citenamefont {Nomura},
  \citenamefont {Corboz}, \citenamefont {Miyata}, \citenamefont {Zherlitsyn},
  \citenamefont {Ishii}, \citenamefont {Kohama}, \citenamefont {Matsuda},
  \citenamefont {Ikeda}, \citenamefont {Zhong}, \citenamefont {Kageyama},\ and\
  \citenamefont {Mila}}]{nomura23}%
  \BibitemOpen
  \bibfield  {author} {\bibinfo {author} {\bibfnamefont {T.}~\bibnamefont
  {Nomura}}, \bibinfo {author} {\bibfnamefont {P.}~\bibnamefont {Corboz}},
  \bibinfo {author} {\bibfnamefont {A.}~\bibnamefont {Miyata}}, \bibinfo
  {author} {\bibfnamefont {S.}~\bibnamefont {Zherlitsyn}}, \bibinfo {author}
  {\bibfnamefont {Y.}~\bibnamefont {Ishii}}, \bibinfo {author} {\bibfnamefont
  {Y.}~\bibnamefont {Kohama}}, \bibinfo {author} {\bibfnamefont {Y.~H.}\
  \bibnamefont {Matsuda}}, \bibinfo {author} {\bibfnamefont {A.}~\bibnamefont
  {Ikeda}}, \bibinfo {author} {\bibfnamefont {C.}~\bibnamefont {Zhong}},
  \bibinfo {author} {\bibfnamefont {H.}~\bibnamefont {Kageyama}},\ and\
  \bibinfo {author} {\bibfnamefont {F.}~\bibnamefont {Mila}},\ }\href
  {https://doi.org/10.1038/s41467-023-39502-5} {\bibfield  {journal} {\bibinfo
  {journal} {Nat Commun}\ }\textbf {\bibinfo {volume} {14}},\ \bibinfo {pages}
  {3769} (\bibinfo {year} {2023})}\BibitemShut {NoStop}%
\bibitem [{\citenamefont {Romh{\'a}nyi}\ \emph {et~al.}(2015)\citenamefont
  {Romh{\'a}nyi}, \citenamefont {Penc},\ and\ \citenamefont
  {Ganesh}}]{romhanyi15}%
  \BibitemOpen
  \bibfield  {author} {\bibinfo {author} {\bibfnamefont {J.}~\bibnamefont
  {Romh{\'a}nyi}}, \bibinfo {author} {\bibfnamefont {K.}~\bibnamefont {Penc}},\
  and\ \bibinfo {author} {\bibfnamefont {R.}~\bibnamefont {Ganesh}},\ }\href
  {https://doi.org/10.1038/ncomms7805} {\bibfield  {journal} {\bibinfo
  {journal} {Nat Commun}\ }\textbf {\bibinfo {volume} {6}},\ \bibinfo {pages}
  {1} (\bibinfo {year} {2015})}\BibitemShut {NoStop}%
\bibitem [{\citenamefont {McClarty}\ \emph {et~al.}(2017)\citenamefont
  {McClarty}, \citenamefont {Kr{\"u}ger}, \citenamefont {Guidi}, \citenamefont
  {Parker}, \citenamefont {Refson}, \citenamefont {Parker}, \citenamefont
  {Prabhakaran},\ and\ \citenamefont {Coldea}}]{mcclarty17}%
  \BibitemOpen
  \bibfield  {author} {\bibinfo {author} {\bibfnamefont {P.~A.}\ \bibnamefont
  {McClarty}}, \bibinfo {author} {\bibfnamefont {F.}~\bibnamefont
  {Kr{\"u}ger}}, \bibinfo {author} {\bibfnamefont {T.}~\bibnamefont {Guidi}},
  \bibinfo {author} {\bibfnamefont {S.~F.}\ \bibnamefont {Parker}}, \bibinfo
  {author} {\bibfnamefont {K.}~\bibnamefont {Refson}}, \bibinfo {author}
  {\bibfnamefont {A.~W.}\ \bibnamefont {Parker}}, \bibinfo {author}
  {\bibfnamefont {D.}~\bibnamefont {Prabhakaran}},\ and\ \bibinfo {author}
  {\bibfnamefont {R.}~\bibnamefont {Coldea}},\ }\href
  {https://doi.org/10.1038/nphys4117} {\bibfield  {journal} {\bibinfo
  {journal} {Nature Phys}\ }\textbf {\bibinfo {volume} {13}},\ \bibinfo {pages}
  {736} (\bibinfo {year} {2017})}\BibitemShut {NoStop}%
\bibitem [{\citenamefont {Jim{\'e}nez}\ \emph {et~al.}(2021)\citenamefont
  {Jim{\'e}nez}, \citenamefont {Crone}, \citenamefont {Fogh}, \citenamefont
  {Zayed}, \citenamefont {Lortz}, \citenamefont {Pomjakushina}, \citenamefont
  {Conder}, \citenamefont {L{\"a}uchli}, \citenamefont {Weber}, \citenamefont
  {Wessel}, \citenamefont {Honecker}, \citenamefont {Normand}, \citenamefont
  {R{\"u}egg}, \citenamefont {Corboz}, \citenamefont {R{\o}nnow},\ and\
  \citenamefont {Mila}}]{jimenez21}%
  \BibitemOpen
  \bibfield  {author} {\bibinfo {author} {\bibfnamefont {J.~L.}\ \bibnamefont
  {Jim{\'e}nez}}, \bibinfo {author} {\bibfnamefont {S.~P.~G.}\ \bibnamefont
  {Crone}}, \bibinfo {author} {\bibfnamefont {E.}~\bibnamefont {Fogh}},
  \bibinfo {author} {\bibfnamefont {M.~E.}\ \bibnamefont {Zayed}}, \bibinfo
  {author} {\bibfnamefont {R.}~\bibnamefont {Lortz}}, \bibinfo {author}
  {\bibfnamefont {E.}~\bibnamefont {Pomjakushina}}, \bibinfo {author}
  {\bibfnamefont {K.}~\bibnamefont {Conder}}, \bibinfo {author} {\bibfnamefont
  {A.~M.}\ \bibnamefont {L{\"a}uchli}}, \bibinfo {author} {\bibfnamefont
  {L.}~\bibnamefont {Weber}}, \bibinfo {author} {\bibfnamefont
  {S.}~\bibnamefont {Wessel}}, \bibinfo {author} {\bibfnamefont
  {A.}~\bibnamefont {Honecker}}, \bibinfo {author} {\bibfnamefont
  {B.}~\bibnamefont {Normand}}, \bibinfo {author} {\bibfnamefont
  {C.}~\bibnamefont {R{\"u}egg}}, \bibinfo {author} {\bibfnamefont
  {P.}~\bibnamefont {Corboz}}, \bibinfo {author} {\bibfnamefont {H.~M.}\
  \bibnamefont {R{\o}nnow}},\ and\ \bibinfo {author} {\bibfnamefont
  {F.}~\bibnamefont {Mila}},\ }\href
  {https://doi.org/10.1038/s41586-021-03411-8} {\bibfield  {journal} {\bibinfo
  {journal} {Nature}\ }\textbf {\bibinfo {volume} {592}},\ \bibinfo {pages}
  {370} (\bibinfo {year} {2021})}\BibitemShut {NoStop}%
\bibitem [{\citenamefont {Cui}\ \emph {et~al.}(2023)\citenamefont {Cui},
  \citenamefont {Liu}, \citenamefont {Lin}, \citenamefont {Wu}, \citenamefont
  {Hong}, \citenamefont {Liu}, \citenamefont {Li}, \citenamefont {Hu},
  \citenamefont {Xi}, \citenamefont {Li}, \citenamefont {Yu}, \citenamefont
  {Sandvik},\ and\ \citenamefont {Yu}}]{cui23}%
  \BibitemOpen
  \bibfield  {author} {\bibinfo {author} {\bibfnamefont {Y.}~\bibnamefont
  {Cui}}, \bibinfo {author} {\bibfnamefont {L.}~\bibnamefont {Liu}}, \bibinfo
  {author} {\bibfnamefont {H.}~\bibnamefont {Lin}}, \bibinfo {author}
  {\bibfnamefont {K.-H.}\ \bibnamefont {Wu}}, \bibinfo {author} {\bibfnamefont
  {W.}~\bibnamefont {Hong}}, \bibinfo {author} {\bibfnamefont {X.}~\bibnamefont
  {Liu}}, \bibinfo {author} {\bibfnamefont {C.}~\bibnamefont {Li}}, \bibinfo
  {author} {\bibfnamefont {Z.}~\bibnamefont {Hu}}, \bibinfo {author}
  {\bibfnamefont {N.}~\bibnamefont {Xi}}, \bibinfo {author} {\bibfnamefont
  {S.}~\bibnamefont {Li}}, \bibinfo {author} {\bibfnamefont {R.}~\bibnamefont
  {Yu}}, \bibinfo {author} {\bibfnamefont {A.~W.}\ \bibnamefont {Sandvik}},\
  and\ \bibinfo {author} {\bibfnamefont {W.}~\bibnamefont {Yu}},\ }\href
  {https://doi.org/10.1126/science.adc9487} {\bibfield  {journal} {\bibinfo
  {journal} {Science}\ }\textbf {\bibinfo {volume} {380}},\ \bibinfo {pages}
  {1179} (\bibinfo {year} {2023})}\BibitemShut {NoStop}%
\bibitem [{\citenamefont {Waki}\ \emph {et~al.}(2007)\citenamefont {Waki},
  \citenamefont {Arai}, \citenamefont {Takigawa}, \citenamefont {Saiga},
  \citenamefont {Uwatoko}, \citenamefont {Kageyama},\ and\ \citenamefont
  {Ueda}}]{waki07}%
  \BibitemOpen
  \bibfield  {author} {\bibinfo {author} {\bibfnamefont {T.}~\bibnamefont
  {Waki}}, \bibinfo {author} {\bibfnamefont {K.}~\bibnamefont {Arai}}, \bibinfo
  {author} {\bibfnamefont {M.}~\bibnamefont {Takigawa}}, \bibinfo {author}
  {\bibfnamefont {Y.}~\bibnamefont {Saiga}}, \bibinfo {author} {\bibfnamefont
  {Y.}~\bibnamefont {Uwatoko}}, \bibinfo {author} {\bibfnamefont
  {H.}~\bibnamefont {Kageyama}},\ and\ \bibinfo {author} {\bibfnamefont
  {Y.}~\bibnamefont {Ueda}},\ }\href {https://doi.org/10.1143/JPSJ.76.073710}
  {\bibfield  {journal} {\bibinfo  {journal} {J. Phys. Soc. Jpn.}\ }\textbf
  {\bibinfo {volume} {76}},\ \bibinfo {pages} {073710} (\bibinfo {year}
  {2007})}\BibitemShut {NoStop}%
\bibitem [{\citenamefont {Haravifard}\ \emph {et~al.}(2016)\citenamefont
  {Haravifard}, \citenamefont {Graf}, \citenamefont {Feiguin}, \citenamefont
  {Batista}, \citenamefont {Lang}, \citenamefont {Silevitch}, \citenamefont
  {Srajer}, \citenamefont {Gaulin}, \citenamefont {Dabkowska},\ and\
  \citenamefont {Rosenbaum}}]{haravifard16}%
  \BibitemOpen
  \bibfield  {author} {\bibinfo {author} {\bibfnamefont {S.}~\bibnamefont
  {Haravifard}}, \bibinfo {author} {\bibfnamefont {D.}~\bibnamefont {Graf}},
  \bibinfo {author} {\bibfnamefont {A.~E.}\ \bibnamefont {Feiguin}}, \bibinfo
  {author} {\bibfnamefont {C.~D.}\ \bibnamefont {Batista}}, \bibinfo {author}
  {\bibfnamefont {J.~C.}\ \bibnamefont {Lang}}, \bibinfo {author}
  {\bibfnamefont {D.~M.}\ \bibnamefont {Silevitch}}, \bibinfo {author}
  {\bibfnamefont {G.}~\bibnamefont {Srajer}}, \bibinfo {author} {\bibfnamefont
  {B.~D.}\ \bibnamefont {Gaulin}}, \bibinfo {author} {\bibfnamefont {H.~A.}\
  \bibnamefont {Dabkowska}},\ and\ \bibinfo {author} {\bibfnamefont {T.~F.}\
  \bibnamefont {Rosenbaum}},\ }\href {https://doi.org/10.1038/ncomms11956}
  {\bibfield  {journal} {\bibinfo  {journal} {Nat. Comm.}\ }\textbf {\bibinfo
  {volume} {7}},\ \bibinfo {pages} {11956} (\bibinfo {year}
  {2016})}\BibitemShut {NoStop}%
\bibitem [{\citenamefont {Zayed}\ \emph {et~al.}(2017)\citenamefont {Zayed},
  \citenamefont {R{\"u}egg}, \citenamefont {J}, \citenamefont {L{\"a}uchli},
  \citenamefont {Panagopoulos}, \citenamefont {Saxena}, \citenamefont
  {Ellerby}, \citenamefont {McMorrow}, \citenamefont {Str{\"a}ssle},
  \citenamefont {Klotz}, \citenamefont {Hamel}, \citenamefont {Sadykov},
  \citenamefont {Pomjakushin}, \citenamefont {Boehm}, \citenamefont
  {Jim{\'e}nez{\textendash}Ruiz}, \citenamefont {Schneidewind}, \citenamefont
  {Pomjakushina}, \citenamefont {Stingaciu}, \citenamefont {Conder},\ and\
  \citenamefont {R{\o}nnow}}]{Zayed17}%
  \BibitemOpen
  \bibfield  {author} {\bibinfo {author} {\bibfnamefont {M.~E.}\ \bibnamefont
  {Zayed}}, \bibinfo {author} {\bibfnamefont {C.}~\bibnamefont {R{\"u}egg}},
  \bibinfo {author} {\bibfnamefont {J.~L.}\ \bibnamefont {J}}, \bibinfo
  {author} {\bibfnamefont {A.~M.}\ \bibnamefont {L{\"a}uchli}}, \bibinfo
  {author} {\bibfnamefont {C.}~\bibnamefont {Panagopoulos}}, \bibinfo {author}
  {\bibfnamefont {S.~S.}\ \bibnamefont {Saxena}}, \bibinfo {author}
  {\bibfnamefont {M.}~\bibnamefont {Ellerby}}, \bibinfo {author} {\bibfnamefont
  {D.~F.}\ \bibnamefont {McMorrow}}, \bibinfo {author} {\bibfnamefont
  {T.}~\bibnamefont {Str{\"a}ssle}}, \bibinfo {author} {\bibfnamefont
  {S.}~\bibnamefont {Klotz}}, \bibinfo {author} {\bibfnamefont
  {G.}~\bibnamefont {Hamel}}, \bibinfo {author} {\bibfnamefont {R.~A.}\
  \bibnamefont {Sadykov}}, \bibinfo {author} {\bibfnamefont {V.}~\bibnamefont
  {Pomjakushin}}, \bibinfo {author} {\bibfnamefont {M.}~\bibnamefont {Boehm}},
  \bibinfo {author} {\bibfnamefont {M.}~\bibnamefont
  {Jim{\'e}nez{\textendash}Ruiz}}, \bibinfo {author} {\bibfnamefont
  {A.}~\bibnamefont {Schneidewind}}, \bibinfo {author} {\bibfnamefont
  {E.}~\bibnamefont {Pomjakushina}}, \bibinfo {author} {\bibfnamefont
  {M.}~\bibnamefont {Stingaciu}}, \bibinfo {author} {\bibfnamefont
  {K.}~\bibnamefont {Conder}},\ and\ \bibinfo {author} {\bibfnamefont {H.~M.}\
  \bibnamefont {R{\o}nnow}},\ }\href {https://doi.org/10.1038/nphys4190}
  {\bibfield  {journal} {\bibinfo  {journal} {Nature Physics}\ }\textbf
  {\bibinfo {volume} {13}},\ \bibinfo {pages} {962} (\bibinfo {year}
  {2017})}\BibitemShut {NoStop}%
\bibitem [{\citenamefont {Sakurai}\ \emph {et~al.}(2018)\citenamefont
  {Sakurai}, \citenamefont {Hirao}, \citenamefont {Hijii}, \citenamefont
  {Okubo}, \citenamefont {Ohta}, \citenamefont {Uwatoko}, \citenamefont
  {Kudo},\ and\ \citenamefont {Koike}}]{sakurai18}%
  \BibitemOpen
  \bibfield  {author} {\bibinfo {author} {\bibfnamefont {T.}~\bibnamefont
  {Sakurai}}, \bibinfo {author} {\bibfnamefont {Y.}~\bibnamefont {Hirao}},
  \bibinfo {author} {\bibfnamefont {K.}~\bibnamefont {Hijii}}, \bibinfo
  {author} {\bibfnamefont {S.}~\bibnamefont {Okubo}}, \bibinfo {author}
  {\bibfnamefont {H.}~\bibnamefont {Ohta}}, \bibinfo {author} {\bibfnamefont
  {Y.}~\bibnamefont {Uwatoko}}, \bibinfo {author} {\bibfnamefont
  {K.}~\bibnamefont {Kudo}},\ and\ \bibinfo {author} {\bibfnamefont
  {Y.}~\bibnamefont {Koike}},\ }\href {https://doi.org/10.7566/JPSJ.87.033701}
  {\bibfield  {journal} {\bibinfo  {journal} {J. Phys. Soc. Jpn.}\ }\textbf
  {\bibinfo {volume} {87}},\ \bibinfo {pages} {033701} (\bibinfo {year}
  {2018})}\BibitemShut {NoStop}%
\bibitem [{\citenamefont {Bettler}\ \emph {et~al.}(2020)\citenamefont
  {Bettler}, \citenamefont {Stoppel}, \citenamefont {Yan}, \citenamefont
  {Gvasaliya},\ and\ \citenamefont {Zheludev}}]{bettler20}%
  \BibitemOpen
  \bibfield  {author} {\bibinfo {author} {\bibfnamefont {S.}~\bibnamefont
  {Bettler}}, \bibinfo {author} {\bibfnamefont {L.}~\bibnamefont {Stoppel}},
  \bibinfo {author} {\bibfnamefont {Z.}~\bibnamefont {Yan}}, \bibinfo {author}
  {\bibfnamefont {S.}~\bibnamefont {Gvasaliya}},\ and\ \bibinfo {author}
  {\bibfnamefont {A.}~\bibnamefont {Zheludev}},\ }\href
  {https://doi.org/10.1103/PhysRevResearch.2.012010} {\bibfield  {journal}
  {\bibinfo  {journal} {Phys. Rev. Research}\ }\textbf {\bibinfo {volume}
  {2}},\ \bibinfo {pages} {012010} (\bibinfo {year} {2020})}\BibitemShut
  {NoStop}%
\bibitem [{\citenamefont {Guo}\ \emph {et~al.}(2020)\citenamefont {Guo},
  \citenamefont {Sun}, \citenamefont {Zhao}, \citenamefont {Wang},
  \citenamefont {Hong}, \citenamefont {Sidorov}, \citenamefont {Ma},
  \citenamefont {Wu}, \citenamefont {Li}, \citenamefont {Meng}, \citenamefont
  {Sandvik},\ and\ \citenamefont {Sun}}]{guo20}%
  \BibitemOpen
  \bibfield  {author} {\bibinfo {author} {\bibfnamefont {J.}~\bibnamefont
  {Guo}}, \bibinfo {author} {\bibfnamefont {G.}~\bibnamefont {Sun}}, \bibinfo
  {author} {\bibfnamefont {B.}~\bibnamefont {Zhao}}, \bibinfo {author}
  {\bibfnamefont {L.}~\bibnamefont {Wang}}, \bibinfo {author} {\bibfnamefont
  {W.}~\bibnamefont {Hong}}, \bibinfo {author} {\bibfnamefont {V.~A.}\
  \bibnamefont {Sidorov}}, \bibinfo {author} {\bibfnamefont {N.}~\bibnamefont
  {Ma}}, \bibinfo {author} {\bibfnamefont {Q.}~\bibnamefont {Wu}}, \bibinfo
  {author} {\bibfnamefont {S.}~\bibnamefont {Li}}, \bibinfo {author}
  {\bibfnamefont {Z.~Y.}\ \bibnamefont {Meng}}, \bibinfo {author}
  {\bibfnamefont {A.~W.}\ \bibnamefont {Sandvik}},\ and\ \bibinfo {author}
  {\bibfnamefont {L.}~\bibnamefont {Sun}},\ }\href
  {https://doi.org/10.1103/PhysRevLett.124.206602} {\bibfield  {journal}
  {\bibinfo  {journal} {Phys. Rev. Lett.}\ }\textbf {\bibinfo {volume} {124}},\
  \bibinfo {pages} {206602} (\bibinfo {year} {2020})}\BibitemShut {NoStop}%
\bibitem [{\citenamefont {Guo}\ \emph {et~al.}(2025)\citenamefont {Guo},
  \citenamefont {Wang}, \citenamefont {Huang}, \citenamefont {Chen},
  \citenamefont {Hong}, \citenamefont {Cai}, \citenamefont {Zhao},
  \citenamefont {Han}, \citenamefont {Chen}, \citenamefont {Zhou},
  \citenamefont {Li}, \citenamefont {Wu}, \citenamefont {Meng},\ and\
  \citenamefont {Sun}}]{guo25}%
  \BibitemOpen
  \bibfield  {author} {\bibinfo {author} {\bibfnamefont {J.}~\bibnamefont
  {Guo}}, \bibinfo {author} {\bibfnamefont {P.}~\bibnamefont {Wang}}, \bibinfo
  {author} {\bibfnamefont {C.}~\bibnamefont {Huang}}, \bibinfo {author}
  {\bibfnamefont {B.-B.}\ \bibnamefont {Chen}}, \bibinfo {author}
  {\bibfnamefont {W.}~\bibnamefont {Hong}}, \bibinfo {author} {\bibfnamefont
  {S.}~\bibnamefont {Cai}}, \bibinfo {author} {\bibfnamefont {J.}~\bibnamefont
  {Zhao}}, \bibinfo {author} {\bibfnamefont {J.}~\bibnamefont {Han}}, \bibinfo
  {author} {\bibfnamefont {X.}~\bibnamefont {Chen}}, \bibinfo {author}
  {\bibfnamefont {Y.}~\bibnamefont {Zhou}}, \bibinfo {author} {\bibfnamefont
  {S.}~\bibnamefont {Li}}, \bibinfo {author} {\bibfnamefont {Q.}~\bibnamefont
  {Wu}}, \bibinfo {author} {\bibfnamefont {Z.~Y.}\ \bibnamefont {Meng}},\ and\
  \bibinfo {author} {\bibfnamefont {L.}~\bibnamefont {Sun}},\ }\href
  {https://doi.org/10.1038/s42005-025-01976-8} {\bibfield  {journal} {\bibinfo
  {journal} {Commun Phys}\ }\textbf {\bibinfo {volume} {8}},\ \bibinfo {pages}
  {1} (\bibinfo {year} {2025})}\BibitemShut {NoStop}%
\bibitem [{\citenamefont {Sriram~Shastry}\ and\ \citenamefont
  {Sutherland}(1981)}]{Shastry81}%
  \BibitemOpen
  \bibfield  {author} {\bibinfo {author} {\bibfnamefont {B.}~\bibnamefont
  {Sriram~Shastry}}\ and\ \bibinfo {author} {\bibfnamefont {B.}~\bibnamefont
  {Sutherland}},\ }\href {https://doi.org/10.1016/0378-4363(81)90838-X}
  {\bibfield  {journal} {\bibinfo  {journal} {Physica {B+C}}\ }\textbf
  {\bibinfo {volume} {108}},\ \bibinfo {pages} {1069} (\bibinfo {year}
  {1981})}\BibitemShut {NoStop}%
\bibitem [{\citenamefont {Miyahara}\ and\ \citenamefont
  {Ueda}(1999)}]{Miyahara99}%
  \BibitemOpen
  \bibfield  {author} {\bibinfo {author} {\bibfnamefont {S.}~\bibnamefont
  {Miyahara}}\ and\ \bibinfo {author} {\bibfnamefont {K.}~\bibnamefont
  {Ueda}},\ }\href {https://doi.org/10.1103/PhysRevLett.82.3701} {\bibfield
  {journal} {\bibinfo  {journal} {Phys. Rev. Lett.}\ }\textbf {\bibinfo
  {volume} {82}},\ \bibinfo {pages} {3701} (\bibinfo {year}
  {1999})}\BibitemShut {NoStop}%
\bibitem [{\citenamefont {Miyahara}\ and\ \citenamefont
  {Ueda}(2003)}]{Miyahara03}%
  \BibitemOpen
  \bibfield  {author} {\bibinfo {author} {\bibfnamefont {S.}~\bibnamefont
  {Miyahara}}\ and\ \bibinfo {author} {\bibfnamefont {K.}~\bibnamefont
  {Ueda}},\ }\href {https://doi.org/10.1088/0953-8984/15/9/201} {\bibfield
  {journal} {\bibinfo  {journal} {J. Phys.: Condensed Matter}\ }\textbf
  {\bibinfo {volume} {15}},\ \bibinfo {pages} {R327} (\bibinfo {year}
  {2003})}\BibitemShut {NoStop}%
\bibitem [{\citenamefont {Corboz}\ and\ \citenamefont
  {Mila}(2014)}]{corboz14_shastry}%
  \BibitemOpen
  \bibfield  {author} {\bibinfo {author} {\bibfnamefont {P.}~\bibnamefont
  {Corboz}}\ and\ \bibinfo {author} {\bibfnamefont {F.}~\bibnamefont {Mila}},\
  }\href {https://doi.org/10.1103/PhysRevLett.112.147203} {\bibfield  {journal}
  {\bibinfo  {journal} {Phys. Rev. Lett.}\ }\textbf {\bibinfo {volume} {112}},\
  \bibinfo {pages} {147203} (\bibinfo {year} {2014})}\BibitemShut {NoStop}%
\bibitem [{\citenamefont {Wessel}\ \emph {et~al.}(2018)\citenamefont {Wessel},
  \citenamefont {Niesen}, \citenamefont {Stapmanns}, \citenamefont {Normand},
  \citenamefont {Mila}, \citenamefont {Corboz},\ and\ \citenamefont
  {Honecker}}]{wessel18}%
  \BibitemOpen
  \bibfield  {author} {\bibinfo {author} {\bibfnamefont {S.}~\bibnamefont
  {Wessel}}, \bibinfo {author} {\bibfnamefont {I.}~\bibnamefont {Niesen}},
  \bibinfo {author} {\bibfnamefont {J.}~\bibnamefont {Stapmanns}}, \bibinfo
  {author} {\bibfnamefont {B.}~\bibnamefont {Normand}}, \bibinfo {author}
  {\bibfnamefont {F.}~\bibnamefont {Mila}}, \bibinfo {author} {\bibfnamefont
  {P.}~\bibnamefont {Corboz}},\ and\ \bibinfo {author} {\bibfnamefont
  {A.}~\bibnamefont {Honecker}},\ }\href
  {https://doi.org/10.1103/PhysRevB.98.174432} {\bibfield  {journal} {\bibinfo
  {journal} {Phys. Rev. B}\ }\textbf {\bibinfo {volume} {98}},\ \bibinfo
  {pages} {174432} (\bibinfo {year} {2018})}\BibitemShut {NoStop}%
\bibitem [{\citenamefont {Wietek}\ \emph {et~al.}(2019)\citenamefont {Wietek},
  \citenamefont {Corboz}, \citenamefont {Wessel}, \citenamefont {Normand},
  \citenamefont {Mila},\ and\ \citenamefont {Honecker}}]{wietek19}%
  \BibitemOpen
  \bibfield  {author} {\bibinfo {author} {\bibfnamefont {A.}~\bibnamefont
  {Wietek}}, \bibinfo {author} {\bibfnamefont {P.}~\bibnamefont {Corboz}},
  \bibinfo {author} {\bibfnamefont {S.}~\bibnamefont {Wessel}}, \bibinfo
  {author} {\bibfnamefont {B.}~\bibnamefont {Normand}}, \bibinfo {author}
  {\bibfnamefont {F.}~\bibnamefont {Mila}},\ and\ \bibinfo {author}
  {\bibfnamefont {A.}~\bibnamefont {Honecker}},\ }\href
  {https://doi.org/10.1103/PhysRevResearch.1.033038} {\bibfield  {journal}
  {\bibinfo  {journal} {Phys. Rev. Res.}\ }\textbf {\bibinfo {volume} {1}},\
  \bibinfo {pages} {033038} (\bibinfo {year} {2019})}\BibitemShut {NoStop}%
\bibitem [{\citenamefont {Czarnik}\ \emph {et~al.}(2021)\citenamefont
  {Czarnik}, \citenamefont {Rams}, \citenamefont {Corboz},\ and\ \citenamefont
  {Dziarmaga}}]{czarnik21}%
  \BibitemOpen
  \bibfield  {author} {\bibinfo {author} {\bibfnamefont {P.}~\bibnamefont
  {Czarnik}}, \bibinfo {author} {\bibfnamefont {M.~M.}\ \bibnamefont {Rams}},
  \bibinfo {author} {\bibfnamefont {P.}~\bibnamefont {Corboz}},\ and\ \bibinfo
  {author} {\bibfnamefont {J.}~\bibnamefont {Dziarmaga}},\ }\href
  {https://doi.org/10.1103/PhysRevB.103.075113} {\bibfield  {journal} {\bibinfo
   {journal} {Phys. Rev. B}\ }\textbf {\bibinfo {volume} {103}},\ \bibinfo
  {pages} {075113} (\bibinfo {year} {2021})}\BibitemShut {NoStop}%
\bibitem [{\citenamefont {Wang}\ \emph {et~al.}(2023)\citenamefont {Wang},
  \citenamefont {Li}, \citenamefont {Xi}, \citenamefont {Gao}, \citenamefont
  {Yan}, \citenamefont {Li},\ and\ \citenamefont {Su}}]{wang23}%
  \BibitemOpen
  \bibfield  {author} {\bibinfo {author} {\bibfnamefont {J.}~\bibnamefont
  {Wang}}, \bibinfo {author} {\bibfnamefont {H.}~\bibnamefont {Li}}, \bibinfo
  {author} {\bibfnamefont {N.}~\bibnamefont {Xi}}, \bibinfo {author}
  {\bibfnamefont {Y.}~\bibnamefont {Gao}}, \bibinfo {author} {\bibfnamefont
  {Q.-B.}\ \bibnamefont {Yan}}, \bibinfo {author} {\bibfnamefont
  {W.}~\bibnamefont {Li}},\ and\ \bibinfo {author} {\bibfnamefont
  {G.}~\bibnamefont {Su}},\ }\href
  {https://doi.org/10.1103/PhysRevLett.131.116702} {\bibfield  {journal}
  {\bibinfo  {journal} {Phys. Rev. Lett.}\ }\textbf {\bibinfo {volume} {131}},\
  \bibinfo {pages} {116702} (\bibinfo {year} {2023})}\BibitemShut {NoStop}%
\bibitem [{\citenamefont {Wang}\ \emph {et~al.}(2026)\citenamefont {Wang},
  \citenamefont {McClarty}, \citenamefont {Dankova}, \citenamefont {Honecker},\
  and\ \citenamefont {Wietek}}]{wang26}%
  \BibitemOpen
  \bibfield  {author} {\bibinfo {author} {\bibfnamefont {Z.}~\bibnamefont
  {Wang}}, \bibinfo {author} {\bibfnamefont {P.}~\bibnamefont {McClarty}},
  \bibinfo {author} {\bibfnamefont {D.}~\bibnamefont {Dankova}}, \bibinfo
  {author} {\bibfnamefont {A.}~\bibnamefont {Honecker}},\ and\ \bibinfo
  {author} {\bibfnamefont {A.}~\bibnamefont {Wietek}},\ }\href
  {https://doi.org/10.1103/ldwx-2s7w} {\bibfield  {journal} {\bibinfo
  {journal} {Phys. Rev. B}\ }\textbf {\bibinfo {volume} {113}},\ \bibinfo
  {pages} {L041104} (\bibinfo {year} {2026})}\BibitemShut {NoStop}%
\bibitem [{\citenamefont {Nyckees}\ \emph {et~al.}(2025)\citenamefont
  {Nyckees}, \citenamefont {Corboz},\ and\ \citenamefont {Mila}}]{nyckees25}%
  \BibitemOpen
  \bibfield  {author} {\bibinfo {author} {\bibfnamefont {S.}~\bibnamefont
  {Nyckees}}, \bibinfo {author} {\bibfnamefont {P.}~\bibnamefont {Corboz}},\
  and\ \bibinfo {author} {\bibfnamefont {F.}~\bibnamefont {Mila}},\ }\href
  {https://doi.org/10.1103/PhysRevB.111.014428} {\bibfield  {journal} {\bibinfo
   {journal} {Phys. Rev. B}\ }\textbf {\bibinfo {volume} {111}},\ \bibinfo
  {pages} {014428} (\bibinfo {year} {2025})}\BibitemShut {NoStop}%
\bibitem [{\citenamefont {Koga}\ and\ \citenamefont {Kawakami}(2000)}]{Koga00}%
  \BibitemOpen
  \bibfield  {author} {\bibinfo {author} {\bibfnamefont {A.}~\bibnamefont
  {Koga}}\ and\ \bibinfo {author} {\bibfnamefont {N.}~\bibnamefont
  {Kawakami}},\ }\href {https://doi.org/10.1103/PhysRevLett.84.4461} {\bibfield
   {journal} {\bibinfo  {journal} {Phys. Rev. Lett.}\ }\textbf {\bibinfo
  {volume} {84}},\ \bibinfo {pages} {4461} (\bibinfo {year}
  {2000})}\BibitemShut {NoStop}%
\bibitem [{\citenamefont {Takushima}\ \emph {et~al.}(2001)\citenamefont
  {Takushima}, \citenamefont {Koga},\ and\ \citenamefont
  {Kawakami}}]{Takushima01}%
  \BibitemOpen
  \bibfield  {author} {\bibinfo {author} {\bibfnamefont {Y.}~\bibnamefont
  {Takushima}}, \bibinfo {author} {\bibfnamefont {A.}~\bibnamefont {Koga}},\
  and\ \bibinfo {author} {\bibfnamefont {N.}~\bibnamefont {Kawakami}},\ }\href
  {https://doi.org/10.1143/JPSJ.70.1369} {\bibfield  {journal} {\bibinfo
  {journal} {J. Phys. Soc. Jpn.}\ }\textbf {\bibinfo {volume} {70}},\ \bibinfo
  {pages} {1369} (\bibinfo {year} {2001})}\BibitemShut {NoStop}%
\bibitem [{\citenamefont {L{\"a}uchli}\ \emph {et~al.}(2002)\citenamefont
  {L{\"a}uchli}, \citenamefont {Wessel},\ and\ \citenamefont
  {Sigrist}}]{Laeuchli02}%
  \BibitemOpen
  \bibfield  {author} {\bibinfo {author} {\bibfnamefont {A.}~\bibnamefont
  {L{\"a}uchli}}, \bibinfo {author} {\bibfnamefont {S.}~\bibnamefont
  {Wessel}},\ and\ \bibinfo {author} {\bibfnamefont {M.}~\bibnamefont
  {Sigrist}},\ }\href {https://doi.org/10.1103/PhysRevB.66.014401} {\bibfield
  {journal} {\bibinfo  {journal} {Phys. Rev. B}\ }\textbf {\bibinfo {volume}
  {66}},\ \bibinfo {pages} {014401} (\bibinfo {year} {2002})}\BibitemShut
  {NoStop}%
\bibitem [{\citenamefont {Corboz}\ and\ \citenamefont
  {Mila}(2013)}]{Corboz13_shastry}%
  \BibitemOpen
  \bibfield  {author} {\bibinfo {author} {\bibfnamefont {P.}~\bibnamefont
  {Corboz}}\ and\ \bibinfo {author} {\bibfnamefont {F.}~\bibnamefont {Mila}},\
  }\href {https://doi.org/10.1103/PhysRevB.87.115144} {\bibfield  {journal}
  {\bibinfo  {journal} {Phys. Rev. B}\ }\textbf {\bibinfo {volume} {87}},\
  \bibinfo {pages} {115144} (\bibinfo {year} {2013})}\BibitemShut {NoStop}%
\bibitem [{\citenamefont {Yang}\ \emph {et~al.}(2022)\citenamefont {Yang},
  \citenamefont {Sandvik},\ and\ \citenamefont {Wang}}]{yang22}%
  \BibitemOpen
  \bibfield  {author} {\bibinfo {author} {\bibfnamefont {J.}~\bibnamefont
  {Yang}}, \bibinfo {author} {\bibfnamefont {A.~W.}\ \bibnamefont {Sandvik}},\
  and\ \bibinfo {author} {\bibfnamefont {L.}~\bibnamefont {Wang}},\ }\href
  {https://doi.org/10.1103/PhysRevB.105.L060409} {\bibfield  {journal}
  {\bibinfo  {journal} {Phys. Rev. B}\ }\textbf {\bibinfo {volume} {105}},\
  \bibinfo {pages} {L060409} (\bibinfo {year} {2022})}\BibitemShut {NoStop}%
\bibitem [{\citenamefont {Wang}\ \emph {et~al.}(2022)\citenamefont {Wang},
  \citenamefont {Zhang},\ and\ \citenamefont {Sandvik}}]{wang22}%
  \BibitemOpen
  \bibfield  {author} {\bibinfo {author} {\bibfnamefont {L.}~\bibnamefont
  {Wang}}, \bibinfo {author} {\bibfnamefont {Y.}~\bibnamefont {Zhang}},\ and\
  \bibinfo {author} {\bibfnamefont {A.~W.}\ \bibnamefont {Sandvik}},\ }\href
  {https://doi.org/10.1088/0256-307X/39/7/077502} {\bibfield  {journal}
  {\bibinfo  {journal} {Chinese Phys. Lett.}\ }\textbf {\bibinfo {volume}
  {39}},\ \bibinfo {pages} {077502} (\bibinfo {year} {2022})}\BibitemShut
  {NoStop}%
\bibitem [{\citenamefont {Kele{\c s}}\ and\ \citenamefont
  {Zhao}(2022)}]{keles22}%
  \BibitemOpen
  \bibfield  {author} {\bibinfo {author} {\bibfnamefont {A.}~\bibnamefont
  {Kele{\c s}}}\ and\ \bibinfo {author} {\bibfnamefont {E.}~\bibnamefont
  {Zhao}},\ }\href {https://doi.org/10.1103/PhysRevB.105.L041115} {\bibfield
  {journal} {\bibinfo  {journal} {Phys. Rev. B}\ }\textbf {\bibinfo {volume}
  {105}},\ \bibinfo {pages} {L041115} (\bibinfo {year} {2022})}\BibitemShut
  {NoStop}%
\bibitem [{\citenamefont {Viteritti}\ \emph {et~al.}(2025)\citenamefont
  {Viteritti}, \citenamefont {Rende}, \citenamefont {Parola}, \citenamefont
  {Goldt},\ and\ \citenamefont {Becca}}]{viteritti25}%
  \BibitemOpen
  \bibfield  {author} {\bibinfo {author} {\bibfnamefont {L.~L.}\ \bibnamefont
  {Viteritti}}, \bibinfo {author} {\bibfnamefont {R.}~\bibnamefont {Rende}},
  \bibinfo {author} {\bibfnamefont {A.}~\bibnamefont {Parola}}, \bibinfo
  {author} {\bibfnamefont {S.}~\bibnamefont {Goldt}},\ and\ \bibinfo {author}
  {\bibfnamefont {F.}~\bibnamefont {Becca}},\ }\href
  {https://doi.org/10.1103/PhysRevB.111.134411} {\bibfield  {journal} {\bibinfo
   {journal} {Phys. Rev. B}\ }\textbf {\bibinfo {volume} {111}},\ \bibinfo
  {pages} {134411} (\bibinfo {year} {2025})}\BibitemShut {NoStop}%
\bibitem [{\citenamefont {Maity}\ \emph {et~al.}(2024)\citenamefont {Maity},
  \citenamefont {Ferrari}, \citenamefont {Lee}, \citenamefont {Potten},
  \citenamefont {M{\"u}ller}, \citenamefont {Thomale}, \citenamefont
  {Samajdar},\ and\ \citenamefont {Iqbal}}]{maity25}%
  \BibitemOpen
  \bibfield  {author} {\bibinfo {author} {\bibfnamefont {A.}~\bibnamefont
  {Maity}}, \bibinfo {author} {\bibfnamefont {F.}~\bibnamefont {Ferrari}},
  \bibinfo {author} {\bibfnamefont {J.~Y.}\ \bibnamefont {Lee}}, \bibinfo
  {author} {\bibfnamefont {J.}~\bibnamefont {Potten}}, \bibinfo {author}
  {\bibfnamefont {T.}~\bibnamefont {M{\"u}ller}}, \bibinfo {author}
  {\bibfnamefont {R.}~\bibnamefont {Thomale}}, \bibinfo {author} {\bibfnamefont
  {R.}~\bibnamefont {Samajdar}},\ and\ \bibinfo {author} {\bibfnamefont
  {Y.}~\bibnamefont {Iqbal}},\ }\bibfield  {journal} {\bibinfo  {journal}
  {arXiv:2501.00096 [cond-mat]}\ }\href
  {https://doi.org/10.48550/arXiv.2501.00096} {10.48550/arXiv.2501.00096}
  (\bibinfo {year} {2024})\BibitemShut {NoStop}%
\bibitem [{\citenamefont {Qian}\ \emph {et~al.}(2024)\citenamefont {Qian},
  \citenamefont {Lv}, \citenamefont {Lee},\ and\ \citenamefont {Qin}}]{qian24}%
  \BibitemOpen
  \bibfield  {author} {\bibinfo {author} {\bibfnamefont {X.}~\bibnamefont
  {Qian}}, \bibinfo {author} {\bibfnamefont {R.}~\bibnamefont {Lv}}, \bibinfo
  {author} {\bibfnamefont {J.~Y.}\ \bibnamefont {Lee}},\ and\ \bibinfo {author}
  {\bibfnamefont {M.}~\bibnamefont {Qin}},\ }\href
  {https://doi.org/10.48550/arXiv.2411.17452} {\bibinfo {title} {From the
  {Shastry}-{Sutherland} model to the {$J_1$-$J_2$} {Heisenberg} model}}
  (\bibinfo {year} {2024}),\ \bibinfo {note} {arXiv:2411.17452
  [cond-mat]}\BibitemShut {NoStop}%
\bibitem [{\citenamefont {Xi}\ \emph {et~al.}(2023)\citenamefont {Xi},
  \citenamefont {Chen}, \citenamefont {Xie},\ and\ \citenamefont {Yu}}]{xi23}%
  \BibitemOpen
  \bibfield  {author} {\bibinfo {author} {\bibfnamefont {N.}~\bibnamefont
  {Xi}}, \bibinfo {author} {\bibfnamefont {H.}~\bibnamefont {Chen}}, \bibinfo
  {author} {\bibfnamefont {Z.~Y.}\ \bibnamefont {Xie}},\ and\ \bibinfo {author}
  {\bibfnamefont {R.}~\bibnamefont {Yu}},\ }\href
  {https://doi.org/10.1103/PhysRevB.107.L220408} {\bibfield  {journal}
  {\bibinfo  {journal} {Phys. Rev. B}\ }\textbf {\bibinfo {volume} {107}},\
  \bibinfo {pages} {L220408} (\bibinfo {year} {2023})}\BibitemShut {NoStop}%
\bibitem [{\citenamefont {Lee}\ \emph {et~al.}(2019)\citenamefont {Lee},
  \citenamefont {You}, \citenamefont {Sachdev},\ and\ \citenamefont
  {Vishwanath}}]{lee19}%
  \BibitemOpen
  \bibfield  {author} {\bibinfo {author} {\bibfnamefont {J.~Y.}\ \bibnamefont
  {Lee}}, \bibinfo {author} {\bibfnamefont {Y.-Z.}\ \bibnamefont {You}},
  \bibinfo {author} {\bibfnamefont {S.}~\bibnamefont {Sachdev}},\ and\ \bibinfo
  {author} {\bibfnamefont {A.}~\bibnamefont {Vishwanath}},\ }\href
  {https://doi.org/10.1103/PhysRevX.9.041037} {\bibfield  {journal} {\bibinfo
  {journal} {Phys. Rev. X}\ }\textbf {\bibinfo {volume} {9}},\ \bibinfo {pages}
  {041037} (\bibinfo {year} {2019})}\BibitemShut {NoStop}%
\bibitem [{\citenamefont {Liu}\ \emph {et~al.}(2024)\citenamefont {Liu},
  \citenamefont {Zhang}, \citenamefont {Wang}, \citenamefont {Gong},
  \citenamefont {Chen},\ and\ \citenamefont {Gu}}]{liu24}%
  \BibitemOpen
  \bibfield  {author} {\bibinfo {author} {\bibfnamefont {W.-Y.}\ \bibnamefont
  {Liu}}, \bibinfo {author} {\bibfnamefont {X.-T.}\ \bibnamefont {Zhang}},
  \bibinfo {author} {\bibfnamefont {Z.}~\bibnamefont {Wang}}, \bibinfo {author}
  {\bibfnamefont {S.-S.}\ \bibnamefont {Gong}}, \bibinfo {author}
  {\bibfnamefont {W.-Q.}\ \bibnamefont {Chen}},\ and\ \bibinfo {author}
  {\bibfnamefont {Z.-C.}\ \bibnamefont {Gu}},\ }\href
  {https://doi.org/10.1103/PhysRevLett.133.026502} {\bibfield  {journal}
  {\bibinfo  {journal} {Phys. Rev. Lett.}\ }\textbf {\bibinfo {volume} {133}},\
  \bibinfo {pages} {026502} (\bibinfo {year} {2024})}\BibitemShut {NoStop}%
\bibitem [{\citenamefont {Verstraete}\ and\ \citenamefont
  {Cirac}(2004)}]{verstraete2004}%
  \BibitemOpen
  \bibfield  {author} {\bibinfo {author} {\bibfnamefont {F.}~\bibnamefont
  {Verstraete}}\ and\ \bibinfo {author} {\bibfnamefont {J.~I.}\ \bibnamefont
  {Cirac}},\ }\href {http://arxiv.org/abs/cond-mat/0407066} {\bibfield
  {journal} {\bibinfo  {journal} {arXiv:cond-mat/0407066}\ } (\bibinfo {year}
  {2004})}\BibitemShut {NoStop}%
\bibitem [{\citenamefont {Nishio}\ \emph {et~al.}(2004)\citenamefont {Nishio},
  \citenamefont {Maeshima}, \citenamefont {Gendiar},\ and\ \citenamefont
  {Nishino}}]{nishio2004}%
  \BibitemOpen
  \bibfield  {author} {\bibinfo {author} {\bibfnamefont {Y.}~\bibnamefont
  {Nishio}}, \bibinfo {author} {\bibfnamefont {N.}~\bibnamefont {Maeshima}},
  \bibinfo {author} {\bibfnamefont {A.}~\bibnamefont {Gendiar}},\ and\ \bibinfo
  {author} {\bibfnamefont {T.}~\bibnamefont {Nishino}},\ }\href@noop {}
  {\bibfield  {journal} {\bibinfo  {journal} {Preprint}\ } (\bibinfo {year}
  {2004})},\ \Eprint {https://arxiv.org/abs/cond-mat/0401115}
  {arXiv:cond-mat/0401115} \BibitemShut {NoStop}%
\bibitem [{\citenamefont {Jordan}\ \emph {et~al.}(2008)\citenamefont {Jordan},
  \citenamefont {Or\'{u}s}, \citenamefont {Vidal}, \citenamefont {Verstraete},\
  and\ \citenamefont {Cirac}}]{jordan2008}%
  \BibitemOpen
  \bibfield  {author} {\bibinfo {author} {\bibfnamefont {J.}~\bibnamefont
  {Jordan}}, \bibinfo {author} {\bibfnamefont {R.}~\bibnamefont {Or\'{u}s}},
  \bibinfo {author} {\bibfnamefont {G.}~\bibnamefont {Vidal}}, \bibinfo
  {author} {\bibfnamefont {F.}~\bibnamefont {Verstraete}},\ and\ \bibinfo
  {author} {\bibfnamefont {J.~I.}\ \bibnamefont {Cirac}},\ }\href
  {https://doi.org/10.1103/PhysRevLett.101.250602} {\bibfield  {journal}
  {\bibinfo  {journal} {Phys. Rev. Lett.}\ }\textbf {\bibinfo {volume} {101}},\
  \bibinfo {pages} {250602} (\bibinfo {year} {2008})}\BibitemShut {NoStop}%
\bibitem [{\citenamefont {Corboz}\ \emph {et~al.}(2014)\citenamefont {Corboz},
  \citenamefont {Rice},\ and\ \citenamefont {Troyer}}]{corboz14_tJ}%
  \BibitemOpen
  \bibfield  {author} {\bibinfo {author} {\bibfnamefont {P.}~\bibnamefont
  {Corboz}}, \bibinfo {author} {\bibfnamefont {T.~M.}\ \bibnamefont {Rice}},\
  and\ \bibinfo {author} {\bibfnamefont {M.}~\bibnamefont {Troyer}},\ }\href
  {https://doi.org/10.1103/PhysRevLett.113.046402} {\bibfield  {journal}
  {\bibinfo  {journal} {Phys. Rev. Lett.}\ }\textbf {\bibinfo {volume} {113}},\
  \bibinfo {pages} {046402} (\bibinfo {year} {2014})}\BibitemShut {NoStop}%
\bibitem [{\citenamefont {Nataf}\ \emph {et~al.}(2016)\citenamefont {Nataf},
  \citenamefont {Lajk{\'o}}, \citenamefont {Corboz}, \citenamefont
  {L{\"a}uchli}, \citenamefont {Penc},\ and\ \citenamefont {Mila}}]{nataf16}%
  \BibitemOpen
  \bibfield  {author} {\bibinfo {author} {\bibfnamefont {P.}~\bibnamefont
  {Nataf}}, \bibinfo {author} {\bibfnamefont {M.}~\bibnamefont {Lajk{\'o}}},
  \bibinfo {author} {\bibfnamefont {P.}~\bibnamefont {Corboz}}, \bibinfo
  {author} {\bibfnamefont {A.~M.}\ \bibnamefont {L{\"a}uchli}}, \bibinfo
  {author} {\bibfnamefont {K.}~\bibnamefont {Penc}},\ and\ \bibinfo {author}
  {\bibfnamefont {F.}~\bibnamefont {Mila}},\ }\href
  {https://doi.org/10.1103/PhysRevB.93.201113} {\bibfield  {journal} {\bibinfo
  {journal} {Phys. Rev. B}\ }\textbf {\bibinfo {volume} {93}},\ \bibinfo
  {pages} {201113} (\bibinfo {year} {2016})}\BibitemShut {NoStop}%
\bibitem [{\citenamefont {Liao}\ \emph {et~al.}(2017)\citenamefont {Liao},
  \citenamefont {Xie}, \citenamefont {Chen}, \citenamefont {Liu}, \citenamefont
  {Xie}, \citenamefont {Huang}, \citenamefont {Normand},\ and\ \citenamefont
  {Xiang}}]{liao17}%
  \BibitemOpen
  \bibfield  {author} {\bibinfo {author} {\bibfnamefont {H.~J.}\ \bibnamefont
  {Liao}}, \bibinfo {author} {\bibfnamefont {Z.~Y.}\ \bibnamefont {Xie}},
  \bibinfo {author} {\bibfnamefont {J.}~\bibnamefont {Chen}}, \bibinfo {author}
  {\bibfnamefont {Z.~Y.}\ \bibnamefont {Liu}}, \bibinfo {author} {\bibfnamefont
  {H.~D.}\ \bibnamefont {Xie}}, \bibinfo {author} {\bibfnamefont {R.~Z.}\
  \bibnamefont {Huang}}, \bibinfo {author} {\bibfnamefont {B.}~\bibnamefont
  {Normand}},\ and\ \bibinfo {author} {\bibfnamefont {T.}~\bibnamefont
  {Xiang}},\ }\href {https://doi.org/10.1103/PhysRevLett.118.137202} {\bibfield
   {journal} {\bibinfo  {journal} {Phys. Rev. Lett.}\ }\textbf {\bibinfo
  {volume} {118}},\ \bibinfo {pages} {137202} (\bibinfo {year}
  {2017})}\BibitemShut {NoStop}%
\bibitem [{\citenamefont {Niesen}\ and\ \citenamefont
  {Corboz}(2017)}]{niesen17}%
  \BibitemOpen
  \bibfield  {author} {\bibinfo {author} {\bibfnamefont {I.}~\bibnamefont
  {Niesen}}\ and\ \bibinfo {author} {\bibfnamefont {P.}~\bibnamefont
  {Corboz}},\ }\href {https://doi.org/10.1103/PhysRevB.95.180404} {\bibfield
  {journal} {\bibinfo  {journal} {Phys. Rev. B}\ }\textbf {\bibinfo {volume}
  {95}},\ \bibinfo {pages} {180404(R)} (\bibinfo {year} {2017})}\BibitemShut
  {NoStop}%
\bibitem [{\citenamefont {Chen}\ \emph {et~al.}(2018)\citenamefont {Chen},
  \citenamefont {Vanderstraeten}, \citenamefont {Capponi},\ and\ \citenamefont
  {Poilblanc}}]{chen18}%
  \BibitemOpen
  \bibfield  {author} {\bibinfo {author} {\bibfnamefont {J.-Y.}\ \bibnamefont
  {Chen}}, \bibinfo {author} {\bibfnamefont {L.}~\bibnamefont
  {Vanderstraeten}}, \bibinfo {author} {\bibfnamefont {S.}~\bibnamefont
  {Capponi}},\ and\ \bibinfo {author} {\bibfnamefont {D.}~\bibnamefont
  {Poilblanc}},\ }\href {https://doi.org/10.1103/PhysRevB.98.184409} {\bibfield
   {journal} {\bibinfo  {journal} {Phys. Rev. B}\ }\textbf {\bibinfo {volume}
  {98}},\ \bibinfo {pages} {184409} (\bibinfo {year} {2018})}\BibitemShut
  {NoStop}%
\bibitem [{\citenamefont {Lee}\ and\ \citenamefont {Kawashima}(2018)}]{lee18}%
  \BibitemOpen
  \bibfield  {author} {\bibinfo {author} {\bibfnamefont {H.-Y.}\ \bibnamefont
  {Lee}}\ and\ \bibinfo {author} {\bibfnamefont {N.}~\bibnamefont
  {Kawashima}},\ }\href {https://doi.org/10.1103/PhysRevB.97.205123} {\bibfield
   {journal} {\bibinfo  {journal} {Phys. Rev. B}\ }\textbf {\bibinfo {volume}
  {97}},\ \bibinfo {pages} {205123} (\bibinfo {year} {2018})}\BibitemShut
  {NoStop}%
\bibitem [{\citenamefont {Jahromi}\ and\ \citenamefont
  {Or{\'u}s}(2018)}]{jahromi18}%
  \BibitemOpen
  \bibfield  {author} {\bibinfo {author} {\bibfnamefont {S.~S.}\ \bibnamefont
  {Jahromi}}\ and\ \bibinfo {author} {\bibfnamefont {R.}~\bibnamefont
  {Or{\'u}s}},\ }\href {https://doi.org/10.1103/PhysRevB.98.155108} {\bibfield
  {journal} {\bibinfo  {journal} {Phys. Rev. B}\ }\textbf {\bibinfo {volume}
  {98}},\ \bibinfo {pages} {155108} (\bibinfo {year} {2018})}\BibitemShut
  {NoStop}%
\bibitem [{\citenamefont {Niesen}\ and\ \citenamefont
  {Corboz}(2018)}]{niesen18}%
  \BibitemOpen
  \bibfield  {author} {\bibinfo {author} {\bibfnamefont {I.}~\bibnamefont
  {Niesen}}\ and\ \bibinfo {author} {\bibfnamefont {P.}~\bibnamefont
  {Corboz}},\ }\href {https://doi.org/10.1103/PhysRevB.97.245146} {\bibfield
  {journal} {\bibinfo  {journal} {Phys. Rev. B}\ }\textbf {\bibinfo {volume}
  {97}},\ \bibinfo {pages} {245146} (\bibinfo {year} {2018})}\BibitemShut
  {NoStop}%
\bibitem [{\citenamefont {Yamaguchi}\ \emph {et~al.}(2018)\citenamefont
  {Yamaguchi}, \citenamefont {Sasaki}, \citenamefont {Okubo}, \citenamefont
  {Yoshida}, \citenamefont {Kida}, \citenamefont {Hagiwara}, \citenamefont
  {Kono}, \citenamefont {Kittaka}, \citenamefont {Sakakibara}, \citenamefont
  {Takigawa}, \citenamefont {Iwasaki},\ and\ \citenamefont
  {Hosokoshi}}]{yamaguchi18}%
  \BibitemOpen
  \bibfield  {author} {\bibinfo {author} {\bibfnamefont {H.}~\bibnamefont
  {Yamaguchi}}, \bibinfo {author} {\bibfnamefont {Y.}~\bibnamefont {Sasaki}},
  \bibinfo {author} {\bibfnamefont {T.}~\bibnamefont {Okubo}}, \bibinfo
  {author} {\bibfnamefont {M.}~\bibnamefont {Yoshida}}, \bibinfo {author}
  {\bibfnamefont {T.}~\bibnamefont {Kida}}, \bibinfo {author} {\bibfnamefont
  {M.}~\bibnamefont {Hagiwara}}, \bibinfo {author} {\bibfnamefont
  {Y.}~\bibnamefont {Kono}}, \bibinfo {author} {\bibfnamefont {S.}~\bibnamefont
  {Kittaka}}, \bibinfo {author} {\bibfnamefont {T.}~\bibnamefont {Sakakibara}},
  \bibinfo {author} {\bibfnamefont {M.}~\bibnamefont {Takigawa}}, \bibinfo
  {author} {\bibfnamefont {Y.}~\bibnamefont {Iwasaki}},\ and\ \bibinfo {author}
  {\bibfnamefont {Y.}~\bibnamefont {Hosokoshi}},\ }\href
  {https://doi.org/10.1103/PhysRevB.98.094402} {\bibfield  {journal} {\bibinfo
  {journal} {Phys. Rev. B}\ }\textbf {\bibinfo {volume} {98}},\ \bibinfo
  {pages} {094402} (\bibinfo {year} {2018})}\BibitemShut {NoStop}%
\bibitem [{\citenamefont {Kshetrimayum}\ \emph {et~al.}(2019)\citenamefont
  {Kshetrimayum}, \citenamefont {Balz}, \citenamefont {Lake},\ and\
  \citenamefont {Eisert}}]{kshetrimayum19b}%
  \BibitemOpen
  \bibfield  {author} {\bibinfo {author} {\bibfnamefont {A.}~\bibnamefont
  {Kshetrimayum}}, \bibinfo {author} {\bibfnamefont {C.}~\bibnamefont {Balz}},
  \bibinfo {author} {\bibfnamefont {B.}~\bibnamefont {Lake}},\ and\ \bibinfo
  {author} {\bibfnamefont {J.}~\bibnamefont {Eisert}},\ }\href
  {http://arxiv.org/abs/1904.00028} {\bibfield  {journal} {\bibinfo  {journal}
  {arXiv:1904.00028 [cond-mat, physics:quant-ph]}\ } (\bibinfo {year}
  {2019})}\BibitemShut {NoStop}%
\bibitem [{\citenamefont {Chung}\ and\ \citenamefont {Corboz}(2019)}]{chung19}%
  \BibitemOpen
  \bibfield  {author} {\bibinfo {author} {\bibfnamefont {S.~S.}\ \bibnamefont
  {Chung}}\ and\ \bibinfo {author} {\bibfnamefont {P.}~\bibnamefont {Corboz}},\
  }\href {https://doi.org/10.1103/PhysRevB.100.035134} {\bibfield  {journal}
  {\bibinfo  {journal} {Phys. Rev. B}\ }\textbf {\bibinfo {volume} {100}},\
  \bibinfo {pages} {035134} (\bibinfo {year} {2019})}\BibitemShut {NoStop}%
\bibitem [{\citenamefont {Ponsioen}\ \emph {et~al.}(2019)\citenamefont
  {Ponsioen}, \citenamefont {Chung},\ and\ \citenamefont
  {Corboz}}]{ponsioen19}%
  \BibitemOpen
  \bibfield  {author} {\bibinfo {author} {\bibfnamefont {B.}~\bibnamefont
  {Ponsioen}}, \bibinfo {author} {\bibfnamefont {S.~S.}\ \bibnamefont
  {Chung}},\ and\ \bibinfo {author} {\bibfnamefont {P.}~\bibnamefont
  {Corboz}},\ }\href {https://doi.org/10.1103/PhysRevB.100.195141} {\bibfield
  {journal} {\bibinfo  {journal} {Phys. Rev. B}\ }\textbf {\bibinfo {volume}
  {100}},\ \bibinfo {pages} {195141} (\bibinfo {year} {2019})}\BibitemShut
  {NoStop}%
\bibitem [{\citenamefont {Lee}\ \emph {et~al.}(2020)\citenamefont {Lee},
  \citenamefont {Kaneko}, \citenamefont {Chern}, \citenamefont {Okubo},
  \citenamefont {Yamaji}, \citenamefont {Kawashima},\ and\ \citenamefont
  {Kim}}]{lee20}%
  \BibitemOpen
  \bibfield  {author} {\bibinfo {author} {\bibfnamefont {H.-Y.}\ \bibnamefont
  {Lee}}, \bibinfo {author} {\bibfnamefont {R.}~\bibnamefont {Kaneko}},
  \bibinfo {author} {\bibfnamefont {L.~E.}\ \bibnamefont {Chern}}, \bibinfo
  {author} {\bibfnamefont {T.}~\bibnamefont {Okubo}}, \bibinfo {author}
  {\bibfnamefont {Y.}~\bibnamefont {Yamaji}}, \bibinfo {author} {\bibfnamefont
  {N.}~\bibnamefont {Kawashima}},\ and\ \bibinfo {author} {\bibfnamefont
  {Y.~B.}\ \bibnamefont {Kim}},\ }\href
  {https://doi.org/10.1038/s41467-020-15320-x} {\bibfield  {journal} {\bibinfo
  {journal} {Nat. Comm.}\ }\textbf {\bibinfo {volume} {11}},\ \bibinfo {pages}
  {1639} (\bibinfo {year} {2020})}\BibitemShut {NoStop}%
\bibitem [{\citenamefont {Gauth{\'e}}\ \emph {et~al.}(2020)\citenamefont
  {Gauth{\'e}}, \citenamefont {Capponi}, \citenamefont {Mambrini},\ and\
  \citenamefont {Poilblanc}}]{gauthe20}%
  \BibitemOpen
  \bibfield  {author} {\bibinfo {author} {\bibfnamefont {O.}~\bibnamefont
  {Gauth{\'e}}}, \bibinfo {author} {\bibfnamefont {S.}~\bibnamefont {Capponi}},
  \bibinfo {author} {\bibfnamefont {M.}~\bibnamefont {Mambrini}},\ and\
  \bibinfo {author} {\bibfnamefont {D.}~\bibnamefont {Poilblanc}},\ }\href
  {https://doi.org/10.1103/PhysRevB.101.205144} {\bibfield  {journal} {\bibinfo
   {journal} {Phys. Rev. B}\ }\textbf {\bibinfo {volume} {101}},\ \bibinfo
  {pages} {205144} (\bibinfo {year} {2020})}\BibitemShut {NoStop}%
\bibitem [{\citenamefont {Hasik}\ \emph {et~al.}(2021)\citenamefont {Hasik},
  \citenamefont {Poilblanc},\ and\ \citenamefont {Becca}}]{hasik21}%
  \BibitemOpen
  \bibfield  {author} {\bibinfo {author} {\bibfnamefont {J.}~\bibnamefont
  {Hasik}}, \bibinfo {author} {\bibfnamefont {D.}~\bibnamefont {Poilblanc}},\
  and\ \bibinfo {author} {\bibfnamefont {F.}~\bibnamefont {Becca}},\ }\href
  {https://doi.org/10.21468/SciPostPhys.10.1.012} {\bibfield  {journal}
  {\bibinfo  {journal} {SciPost Physics}\ }\textbf {\bibinfo {volume} {10}},\
  \bibinfo {pages} {012} (\bibinfo {year} {2021})}\BibitemShut {NoStop}%
\bibitem [{\citenamefont {Liu}\ \emph {et~al.}(2022)\citenamefont {Liu},
  \citenamefont {Hasik}, \citenamefont {Gong}, \citenamefont {Poilblanc},
  \citenamefont {Chen},\ and\ \citenamefont {Gu}}]{liu22b}%
  \BibitemOpen
  \bibfield  {author} {\bibinfo {author} {\bibfnamefont {W.-Y.}\ \bibnamefont
  {Liu}}, \bibinfo {author} {\bibfnamefont {J.}~\bibnamefont {Hasik}}, \bibinfo
  {author} {\bibfnamefont {S.-S.}\ \bibnamefont {Gong}}, \bibinfo {author}
  {\bibfnamefont {D.}~\bibnamefont {Poilblanc}}, \bibinfo {author}
  {\bibfnamefont {W.-Q.}\ \bibnamefont {Chen}},\ and\ \bibinfo {author}
  {\bibfnamefont {Z.-C.}\ \bibnamefont {Gu}},\ }\href
  {https://doi.org/10.1103/PhysRevX.12.031039} {\bibfield  {journal} {\bibinfo
  {journal} {Phys. Rev. X}\ }\textbf {\bibinfo {volume} {12}},\ \bibinfo
  {pages} {031039} (\bibinfo {year} {2022})}\BibitemShut {NoStop}%
\bibitem [{\citenamefont {Peschke}\ \emph {et~al.}(2022)\citenamefont
  {Peschke}, \citenamefont {Ponsioen},\ and\ \citenamefont
  {Corboz}}]{peschke22}%
  \BibitemOpen
  \bibfield  {author} {\bibinfo {author} {\bibfnamefont {M.}~\bibnamefont
  {Peschke}}, \bibinfo {author} {\bibfnamefont {B.}~\bibnamefont {Ponsioen}},\
  and\ \bibinfo {author} {\bibfnamefont {P.}~\bibnamefont {Corboz}},\ }\href
  {https://doi.org/10.1103/PhysRevB.106.205140} {\bibfield  {journal} {\bibinfo
   {journal} {Phys. Rev. B}\ }\textbf {\bibinfo {volume} {106}},\ \bibinfo
  {pages} {205140} (\bibinfo {year} {2022})}\BibitemShut {NoStop}%
\bibitem [{\citenamefont {Hasik}\ \emph {et~al.}(2022)\citenamefont {Hasik},
  \citenamefont {Van~Damme}, \citenamefont {Poilblanc},\ and\ \citenamefont
  {Vanderstraeten}}]{hasik22}%
  \BibitemOpen
  \bibfield  {author} {\bibinfo {author} {\bibfnamefont {J.}~\bibnamefont
  {Hasik}}, \bibinfo {author} {\bibfnamefont {M.}~\bibnamefont {Van~Damme}},
  \bibinfo {author} {\bibfnamefont {D.}~\bibnamefont {Poilblanc}},\ and\
  \bibinfo {author} {\bibfnamefont {L.}~\bibnamefont {Vanderstraeten}},\ }\href
  {https://doi.org/10.1103/PhysRevLett.129.177201} {\bibfield  {journal}
  {\bibinfo  {journal} {Phys. Rev. Lett.}\ }\textbf {\bibinfo {volume} {129}},\
  \bibinfo {pages} {177201} (\bibinfo {year} {2022})}\BibitemShut {NoStop}%
\bibitem [{\citenamefont {Ponsioen}\ \emph {et~al.}(2023)\citenamefont
  {Ponsioen}, \citenamefont {Chung},\ and\ \citenamefont
  {Corboz}}]{ponsioen23b}%
  \BibitemOpen
  \bibfield  {author} {\bibinfo {author} {\bibfnamefont {B.}~\bibnamefont
  {Ponsioen}}, \bibinfo {author} {\bibfnamefont {S.~S.}\ \bibnamefont
  {Chung}},\ and\ \bibinfo {author} {\bibfnamefont {P.}~\bibnamefont
  {Corboz}},\ }\href {https://doi.org/10.1103/PhysRevB.108.205154} {\bibfield
  {journal} {\bibinfo  {journal} {Phys. Rev. B}\ }\textbf {\bibinfo {volume}
  {108}},\ \bibinfo {pages} {205154} (\bibinfo {year} {2023})}\BibitemShut
  {NoStop}%
\bibitem [{\citenamefont {Weerda}\ and\ \citenamefont
  {Rizzi}(2024)}]{weerda24}%
  \BibitemOpen
  \bibfield  {author} {\bibinfo {author} {\bibfnamefont {E.~L.}\ \bibnamefont
  {Weerda}}\ and\ \bibinfo {author} {\bibfnamefont {M.}~\bibnamefont {Rizzi}},\
  }\href {https://doi.org/10.1103/PhysRevB.109.L241117} {\bibfield  {journal}
  {\bibinfo  {journal} {Phys. Rev. B}\ }\textbf {\bibinfo {volume} {109}},\
  \bibinfo {pages} {L241117} (\bibinfo {year} {2024})}\BibitemShut {NoStop}%
\bibitem [{\citenamefont {Xu}\ \emph {et~al.}(2023)\citenamefont {Xu},
  \citenamefont {Capponi}, \citenamefont {Chen}, \citenamefont
  {Vanderstraeten}, \citenamefont {Hasik}, \citenamefont {Nevidomskyy},
  \citenamefont {Mambrini}, \citenamefont {Penc},\ and\ \citenamefont
  {Poilblanc}}]{xu23b}%
  \BibitemOpen
  \bibfield  {author} {\bibinfo {author} {\bibfnamefont {Y.}~\bibnamefont
  {Xu}}, \bibinfo {author} {\bibfnamefont {S.}~\bibnamefont {Capponi}},
  \bibinfo {author} {\bibfnamefont {J.-Y.}\ \bibnamefont {Chen}}, \bibinfo
  {author} {\bibfnamefont {L.}~\bibnamefont {Vanderstraeten}}, \bibinfo
  {author} {\bibfnamefont {J.}~\bibnamefont {Hasik}}, \bibinfo {author}
  {\bibfnamefont {A.~H.}\ \bibnamefont {Nevidomskyy}}, \bibinfo {author}
  {\bibfnamefont {M.}~\bibnamefont {Mambrini}}, \bibinfo {author}
  {\bibfnamefont {K.}~\bibnamefont {Penc}},\ and\ \bibinfo {author}
  {\bibfnamefont {D.}~\bibnamefont {Poilblanc}},\ }\href
  {https://doi.org/10.1103/PhysRevB.108.195153} {\bibfield  {journal} {\bibinfo
   {journal} {Phys. Rev. B}\ }\textbf {\bibinfo {volume} {108}},\ \bibinfo
  {pages} {195153} (\bibinfo {year} {2023})}\BibitemShut {NoStop}%
\bibitem [{\citenamefont {Hasik}\ and\ \citenamefont {Corboz}(2024)}]{hasik24}%
  \BibitemOpen
  \bibfield  {author} {\bibinfo {author} {\bibfnamefont {J.}~\bibnamefont
  {Hasik}}\ and\ \bibinfo {author} {\bibfnamefont {P.}~\bibnamefont {Corboz}},\
  }\href {https://doi.org/10.1103/PhysRevLett.133.176502} {\bibfield  {journal}
  {\bibinfo  {journal} {Phys. Rev. Lett.}\ }\textbf {\bibinfo {volume} {133}},\
  \bibinfo {pages} {176502} (\bibinfo {year} {2024})}\BibitemShut {NoStop}%
\bibitem [{\citenamefont {Schmoll}\ \emph {et~al.}(2024)\citenamefont
  {Schmoll}, \citenamefont {Naumann}, \citenamefont {Eisert},\ and\
  \citenamefont {Iqbal}}]{schmoll24}%
  \BibitemOpen
  \bibfield  {author} {\bibinfo {author} {\bibfnamefont {P.}~\bibnamefont
  {Schmoll}}, \bibinfo {author} {\bibfnamefont {J.}~\bibnamefont {Naumann}},
  \bibinfo {author} {\bibfnamefont {J.}~\bibnamefont {Eisert}},\ and\ \bibinfo
  {author} {\bibfnamefont {Y.}~\bibnamefont {Iqbal}},\ }\href
  {http://arxiv.org/abs/2407.07145} {\bibfield  {journal} {\bibinfo  {journal}
  {arXiv:2407.07145 [cond-mat]}\ } (\bibinfo {year} {2024})}\BibitemShut
  {NoStop}%
\bibitem [{\citenamefont {Singh}\ \emph {et~al.}(2011)\citenamefont {Singh},
  \citenamefont {Pfeifer},\ and\ \citenamefont {Vidal}}]{singh2010}%
  \BibitemOpen
  \bibfield  {author} {\bibinfo {author} {\bibfnamefont {S.}~\bibnamefont
  {Singh}}, \bibinfo {author} {\bibfnamefont {R.~N.~C.}\ \bibnamefont
  {Pfeifer}},\ and\ \bibinfo {author} {\bibfnamefont {G.}~\bibnamefont
  {Vidal}},\ }\href {https://doi.org/10.1103/PhysRevB.83.115125} {\bibfield
  {journal} {\bibinfo  {journal} {Phys. Rev. B}\ }\textbf {\bibinfo {volume}
  {83}},\ \bibinfo {pages} {115125} (\bibinfo {year} {2011})}\BibitemShut
  {NoStop}%
\bibitem [{\citenamefont {Bauer}\ \emph {et~al.}(2011)\citenamefont {Bauer},
  \citenamefont {Corboz}, \citenamefont {Or\'us},\ and\ \citenamefont
  {Troyer}}]{bauer2011}%
  \BibitemOpen
  \bibfield  {author} {\bibinfo {author} {\bibfnamefont {B.}~\bibnamefont
  {Bauer}}, \bibinfo {author} {\bibfnamefont {P.}~\bibnamefont {Corboz}},
  \bibinfo {author} {\bibfnamefont {R.}~\bibnamefont {Or\'us}},\ and\ \bibinfo
  {author} {\bibfnamefont {M.}~\bibnamefont {Troyer}},\ }\href
  {https://doi.org/10.1103/PhysRevB.83.125106} {\bibfield  {journal} {\bibinfo
  {journal} {Phys. Rev. B}\ }\textbf {\bibinfo {volume} {83}},\ \bibinfo
  {pages} {125106} (\bibinfo {year} {2011})}\BibitemShut {NoStop}%
\bibitem [{\citenamefont {Nishino}\ and\ \citenamefont
  {Okunishi}(1996)}]{nishino1996}%
  \BibitemOpen
  \bibfield  {author} {\bibinfo {author} {\bibfnamefont {T.}~\bibnamefont
  {Nishino}}\ and\ \bibinfo {author} {\bibfnamefont {K.}~\bibnamefont
  {Okunishi}},\ }\href {https://doi.org/10.1143/JPSJ.65.891} {\bibfield
  {journal} {\bibinfo  {journal} {J. Phys. Soc. Jpn.}\ }\textbf {\bibinfo
  {volume} {65}},\ \bibinfo {pages} {891} (\bibinfo {year} {1996})}\BibitemShut
  {NoStop}%
\bibitem [{\citenamefont {Or\'{u}s}\ \emph {et~al.}(2009)\citenamefont
  {Or\'{u}s}, \citenamefont {Doherty},\ and\ \citenamefont {Vidal}}]{Orus2009}%
  \BibitemOpen
  \bibfield  {author} {\bibinfo {author} {\bibfnamefont {R.}~\bibnamefont
  {Or\'{u}s}}, \bibinfo {author} {\bibfnamefont {A.~C.}\ \bibnamefont
  {Doherty}},\ and\ \bibinfo {author} {\bibfnamefont {G.}~\bibnamefont
  {Vidal}},\ }\href {https://doi.org/10.1103/PhysRevLett.102.077203} {\bibfield
   {journal} {\bibinfo  {journal} {Phys. Rev. Lett.}\ }\textbf {\bibinfo
  {volume} {102}},\ \bibinfo {pages} {077203} (\bibinfo {year}
  {2009})}\BibitemShut {NoStop}%
\bibitem [{\citenamefont {Corboz}\ \emph {et~al.}(2011)\citenamefont {Corboz},
  \citenamefont {White}, \citenamefont {Vidal},\ and\ \citenamefont
  {Troyer}}]{Corboz2011}%
  \BibitemOpen
  \bibfield  {author} {\bibinfo {author} {\bibfnamefont {P.}~\bibnamefont
  {Corboz}}, \bibinfo {author} {\bibfnamefont {S.~R.}\ \bibnamefont {White}},
  \bibinfo {author} {\bibfnamefont {G.}~\bibnamefont {Vidal}},\ and\ \bibinfo
  {author} {\bibfnamefont {M.}~\bibnamefont {Troyer}},\ }\href
  {https://doi.org/10.1103/PhysRevB.84.041108} {\bibfield  {journal} {\bibinfo
  {journal} {Phys. Rev. B}\ }\textbf {\bibinfo {volume} {84}},\ \bibinfo
  {pages} {041108(R)} (\bibinfo {year} {2011})}\BibitemShut {NoStop}%
\bibitem [{\citenamefont {Liao}\ \emph {et~al.}(2019)\citenamefont {Liao},
  \citenamefont {Liu}, \citenamefont {Wang},\ and\ \citenamefont
  {Xiang}}]{liao19}%
  \BibitemOpen
  \bibfield  {author} {\bibinfo {author} {\bibfnamefont {H.-J.}\ \bibnamefont
  {Liao}}, \bibinfo {author} {\bibfnamefont {J.-G.}\ \bibnamefont {Liu}},
  \bibinfo {author} {\bibfnamefont {L.}~\bibnamefont {Wang}},\ and\ \bibinfo
  {author} {\bibfnamefont {T.}~\bibnamefont {Xiang}},\ }\href
  {https://doi.org/10.1103/PhysRevX.9.031041} {\bibfield  {journal} {\bibinfo
  {journal} {Phys. Rev. X}\ }\textbf {\bibinfo {volume} {9}},\ \bibinfo {pages}
  {031041} (\bibinfo {year} {2019})}\BibitemShut {NoStop}%
\bibitem [{\citenamefont {Ponsioen}\ \emph {et~al.}(2022)\citenamefont
  {Ponsioen}, \citenamefont {Assaad},\ and\ \citenamefont
  {Corboz}}]{ponsioen22}%
  \BibitemOpen
  \bibfield  {author} {\bibinfo {author} {\bibfnamefont {B.}~\bibnamefont
  {Ponsioen}}, \bibinfo {author} {\bibfnamefont {F.}~\bibnamefont {Assaad}},\
  and\ \bibinfo {author} {\bibfnamefont {P.}~\bibnamefont {Corboz}},\ }\href
  {https://doi.org/10.21468/SciPostPhys.12.1.006} {\bibfield  {journal}
  {\bibinfo  {journal} {SciPost Physics}\ }\textbf {\bibinfo {volume} {12}},\
  \bibinfo {pages} {006} (\bibinfo {year} {2022})}\BibitemShut {NoStop}%
\bibitem [{\citenamefont {Jiang}\ \emph {et~al.}(2008)\citenamefont {Jiang},
  \citenamefont {Weng},\ and\ \citenamefont {Xiang}}]{jiang2008}%
  \BibitemOpen
  \bibfield  {author} {\bibinfo {author} {\bibfnamefont {H.~C.}\ \bibnamefont
  {Jiang}}, \bibinfo {author} {\bibfnamefont {Z.~Y.}\ \bibnamefont {Weng}},\
  and\ \bibinfo {author} {\bibfnamefont {T.}~\bibnamefont {Xiang}},\ }\href
  {https://doi.org/10.1103/PhysRevLett.101.090603} {\bibfield  {journal}
  {\bibinfo  {journal} {Phys. Rev. Lett.}\ }\textbf {\bibinfo {volume} {101}},\
  \bibinfo {pages} {090603} (\bibinfo {year} {2008})}\BibitemShut {NoStop}%
\bibitem [{\citenamefont {Phien}\ \emph {et~al.}(2015)\citenamefont {Phien},
  \citenamefont {Bengua}, \citenamefont {Tuan}, \citenamefont {Corboz},\ and\
  \citenamefont {Orus}}]{phien15}%
  \BibitemOpen
  \bibfield  {author} {\bibinfo {author} {\bibfnamefont {H.~N.}\ \bibnamefont
  {Phien}}, \bibinfo {author} {\bibfnamefont {J.~A.}\ \bibnamefont {Bengua}},
  \bibinfo {author} {\bibfnamefont {H.~D.}\ \bibnamefont {Tuan}}, \bibinfo
  {author} {\bibfnamefont {P.}~\bibnamefont {Corboz}},\ and\ \bibinfo {author}
  {\bibfnamefont {R.}~\bibnamefont {Orus}},\ }\href
  {https://doi.org/10.1103/PhysRevB.92.035142} {\bibfield  {journal} {\bibinfo
  {journal} {Phys. Rev. B}\ }\textbf {\bibinfo {volume} {92}},\ \bibinfo
  {pages} {035142} (\bibinfo {year} {2015})}\BibitemShut {NoStop}%
\bibitem [{\citenamefont {Corboz}(2016)}]{corboz16b}%
  \BibitemOpen
  \bibfield  {author} {\bibinfo {author} {\bibfnamefont {P.}~\bibnamefont
  {Corboz}},\ }\href {https://doi.org/10.1103/PhysRevB.94.035133} {\bibfield
  {journal} {\bibinfo  {journal} {Phys. Rev. B}\ }\textbf {\bibinfo {volume}
  {94}},\ \bibinfo {pages} {035133} (\bibinfo {year} {2016})}\BibitemShut
  {NoStop}%
\bibitem [{\citenamefont {Francuz}\ \emph {et~al.}(2025)\citenamefont
  {Francuz}, \citenamefont {Schuch},\ and\ \citenamefont
  {Vanhecke}}]{francuz25}%
  \BibitemOpen
  \bibfield  {author} {\bibinfo {author} {\bibfnamefont {A.}~\bibnamefont
  {Francuz}}, \bibinfo {author} {\bibfnamefont {N.}~\bibnamefont {Schuch}},\
  and\ \bibinfo {author} {\bibfnamefont {B.}~\bibnamefont {Vanhecke}},\ }\href
  {https://doi.org/10.1103/PhysRevResearch.7.013237} {\bibfield  {journal}
  {\bibinfo  {journal} {Phys. Rev. Res.}\ }\textbf {\bibinfo {volume} {7}},\
  \bibinfo {pages} {013237} (\bibinfo {year} {2025})}\BibitemShut {NoStop}%
\bibitem [{\citenamefont {Cort{\'e}s~Estay}\ \emph {et~al.}(2025)\citenamefont
  {Cort{\'e}s~Estay}, \citenamefont {Kamar},\ and\ \citenamefont
  {Corboz}}]{cortes25}%
  \BibitemOpen
  \bibfield  {author} {\bibinfo {author} {\bibfnamefont {E.}~\bibnamefont
  {Cort{\'e}s~Estay}}, \bibinfo {author} {\bibfnamefont {N.~A.}\ \bibnamefont
  {Kamar}},\ and\ \bibinfo {author} {\bibfnamefont {P.}~\bibnamefont
  {Corboz}},\ }\href {http://arxiv.org/abs/2511.22669} {\bibfield  {journal}
  {\bibinfo  {journal} {arXiv:2511.22669 [cond-mat]}\ } (\bibinfo {year}
  {2025})}\BibitemShut {NoStop}%
\bibitem [{\citenamefont {Arias~Espinoza}\ and\ \citenamefont
  {Corboz}(2024)}]{arias24}%
  \BibitemOpen
  \bibfield  {author} {\bibinfo {author} {\bibfnamefont {J.~D.}\ \bibnamefont
  {Arias~Espinoza}}\ and\ \bibinfo {author} {\bibfnamefont {P.}~\bibnamefont
  {Corboz}},\ }\href {https://doi.org/10.1103/PhysRevB.110.094314} {\bibfield
  {journal} {\bibinfo  {journal} {Phys. Rev. B}\ }\textbf {\bibinfo {volume}
  {110}},\ \bibinfo {pages} {094314} (\bibinfo {year} {2024})}\BibitemShut
  {NoStop}%
\bibitem [{\citenamefont {Vanderstraeten}\ \emph {et~al.}(2016)\citenamefont
  {Vanderstraeten}, \citenamefont {Haegeman}, \citenamefont {Corboz},\ and\
  \citenamefont {Verstraete}}]{vanderstraeten16}%
  \BibitemOpen
  \bibfield  {author} {\bibinfo {author} {\bibfnamefont {L.}~\bibnamefont
  {Vanderstraeten}}, \bibinfo {author} {\bibfnamefont {J.}~\bibnamefont
  {Haegeman}}, \bibinfo {author} {\bibfnamefont {P.}~\bibnamefont {Corboz}},\
  and\ \bibinfo {author} {\bibfnamefont {F.}~\bibnamefont {Verstraete}},\
  }\href {https://doi.org/10.1103/PhysRevB.94.155123} {\bibfield  {journal}
  {\bibinfo  {journal} {Phys. Rev. B}\ }\textbf {\bibinfo {volume} {94}},\
  \bibinfo {pages} {155123} (\bibinfo {year} {2016})}\BibitemShut {NoStop}%
\bibitem [{Note1()}]{Note1}%
  \BibitemOpen
  \bibinfo {note} {Extrapolations based on the bond dimension are generally not
  very accurate, because the functional behavior of E with respect to D is
  typically not smooth, and the functional form is unknown. In contrast, the
  energy as a function of the variance is much smoother, and close to
  convergence, the energy depends linearly on the variance. Here we have used a
  second-order polynomial fit to include corrections to the linear
  scaling}\BibitemShut {NoStop}%
\bibitem [{Note2()}]{Note2}%
  \BibitemOpen
  \bibinfo {note} {The extrapolations of the energies based on a second order
  polynomial, using the three largest $D$ values for the plaquette state. For
  the AF state we took the average of the extrapolated values based on the
  three and four largest $D$ values}\BibitemShut {NoStop}%
\bibitem [{Note3()}]{Note3}%
  \BibitemOpen
  \bibinfo {note} {A weak first-order transition refers to a discontinuous
  transition that is close to being continuous. It typically exhibits only mild
  hysteresis around the transition point; the discontinuity in the order
  parameter is small, and the energy curves of the two states intersect at a
  shallow angle~\cite {demidio23}}\BibitemShut {NoStop}%
\bibitem [{\citenamefont {Corboz}\ \emph {et~al.}(2018)\citenamefont {Corboz},
  \citenamefont {Czarnik}, \citenamefont {Kapteijns},\ and\ \citenamefont
  {Tagliacozzo}}]{corboz18}%
  \BibitemOpen
  \bibfield  {author} {\bibinfo {author} {\bibfnamefont {P.}~\bibnamefont
  {Corboz}}, \bibinfo {author} {\bibfnamefont {P.}~\bibnamefont {Czarnik}},
  \bibinfo {author} {\bibfnamefont {G.}~\bibnamefont {Kapteijns}},\ and\
  \bibinfo {author} {\bibfnamefont {L.}~\bibnamefont {Tagliacozzo}},\ }\href
  {https://doi.org/10.1103/PhysRevX.8.031031} {\bibfield  {journal} {\bibinfo
  {journal} {Phys. Rev. X}\ }\textbf {\bibinfo {volume} {8}},\ \bibinfo {pages}
  {031031} (\bibinfo {year} {2018})}\BibitemShut {NoStop}%
\bibitem [{\citenamefont {Rader}\ and\ \citenamefont
  {L{\"a}uchli}(2018)}]{rader18}%
  \BibitemOpen
  \bibfield  {author} {\bibinfo {author} {\bibfnamefont {M.}~\bibnamefont
  {Rader}}\ and\ \bibinfo {author} {\bibfnamefont {A.~M.}\ \bibnamefont
  {L{\"a}uchli}},\ }\href {https://doi.org/10.1103/PhysRevX.8.031030}
  {\bibfield  {journal} {\bibinfo  {journal} {Phys. Rev. X}\ }\textbf {\bibinfo
  {volume} {8}},\ \bibinfo {pages} {031030} (\bibinfo {year}
  {2018})}\BibitemShut {NoStop}%
\bibitem [{\citenamefont {Vanhecke}\ \emph {et~al.}(2022)\citenamefont
  {Vanhecke}, \citenamefont {Hasik}, \citenamefont {Verstraete},\ and\
  \citenamefont {Vanderstraeten}}]{vanhecke22}%
  \BibitemOpen
  \bibfield  {author} {\bibinfo {author} {\bibfnamefont {B.}~\bibnamefont
  {Vanhecke}}, \bibinfo {author} {\bibfnamefont {J.}~\bibnamefont {Hasik}},
  \bibinfo {author} {\bibfnamefont {F.}~\bibnamefont {Verstraete}},\ and\
  \bibinfo {author} {\bibfnamefont {L.}~\bibnamefont {Vanderstraeten}},\ }\href
  {https://doi.org/10.1103/PhysRevLett.129.200601} {\bibfield  {journal}
  {\bibinfo  {journal} {Phys. Rev. Lett.}\ }\textbf {\bibinfo {volume} {129}},\
  \bibinfo {pages} {200601} (\bibinfo {year} {2022})}\BibitemShut {NoStop}%
\bibitem [{\citenamefont {Tagliacozzo}\ \emph {et~al.}(2008)\citenamefont
  {Tagliacozzo}, \citenamefont {de~Oliveira}, \citenamefont {Iblisdir},\ and\
  \citenamefont {Latorre}}]{tagliacozzo08}%
  \BibitemOpen
  \bibfield  {author} {\bibinfo {author} {\bibfnamefont {L.}~\bibnamefont
  {Tagliacozzo}}, \bibinfo {author} {\bibfnamefont {T.~R.}\ \bibnamefont
  {de~Oliveira}}, \bibinfo {author} {\bibfnamefont {S.}~\bibnamefont
  {Iblisdir}},\ and\ \bibinfo {author} {\bibfnamefont {J.~I.}\ \bibnamefont
  {Latorre}},\ }\href {https://doi.org/10.1103/PhysRevB.78.024410} {\bibfield
  {journal} {\bibinfo  {journal} {Phys. Rev. B}\ }\textbf {\bibinfo {volume}
  {78}},\ \bibinfo {pages} {024410} (\bibinfo {year} {2008})}\BibitemShut
  {NoStop}%
\bibitem [{\citenamefont {Pollmann}\ \emph {et~al.}(2009)\citenamefont
  {Pollmann}, \citenamefont {Mukerjee}, \citenamefont {Turner},\ and\
  \citenamefont {Moore}}]{pollmann2009}%
  \BibitemOpen
  \bibfield  {author} {\bibinfo {author} {\bibfnamefont {F.}~\bibnamefont
  {Pollmann}}, \bibinfo {author} {\bibfnamefont {S.}~\bibnamefont {Mukerjee}},
  \bibinfo {author} {\bibfnamefont {A.~M.}\ \bibnamefont {Turner}},\ and\
  \bibinfo {author} {\bibfnamefont {J.~E.}\ \bibnamefont {Moore}},\ }\href
  {https://doi.org/10.1103/PhysRevLett.102.255701} {\bibfield  {journal}
  {\bibinfo  {journal} {Phys. Rev. Lett.}\ }\textbf {\bibinfo {volume} {102}},\
  \bibinfo {pages} {255701} (\bibinfo {year} {2009})}\BibitemShut {NoStop}%
\bibitem [{\citenamefont {Hasenfratz}\ and\ \citenamefont
  {Niedermayer}(1993)}]{hasenfratz93}%
  \BibitemOpen
  \bibfield  {author} {\bibinfo {author} {\bibfnamefont {P.}~\bibnamefont
  {Hasenfratz}}\ and\ \bibinfo {author} {\bibfnamefont {F.}~\bibnamefont
  {Niedermayer}},\ }\href {https://doi.org/10.1007/BF01309171} {\bibfield
  {journal} {\bibinfo  {journal} {Z. Physik B - Condensed Matter}\ }\textbf
  {\bibinfo {volume} {92}},\ \bibinfo {pages} {91} (\bibinfo {year}
  {1993})}\BibitemShut {NoStop}%
\bibitem [{\citenamefont {Rams}\ \emph {et~al.}(2018)\citenamefont {Rams},
  \citenamefont {Czarnik},\ and\ \citenamefont {Cincio}}]{rams18}%
  \BibitemOpen
  \bibfield  {author} {\bibinfo {author} {\bibfnamefont {M.~M.}\ \bibnamefont
  {Rams}}, \bibinfo {author} {\bibfnamefont {P.}~\bibnamefont {Czarnik}},\ and\
  \bibinfo {author} {\bibfnamefont {L.}~\bibnamefont {Cincio}},\ }\href
  {https://doi.org/10.1103/PhysRevX.8.041033} {\bibfield  {journal} {\bibinfo
  {journal} {Phys. Rev. X}\ }\textbf {\bibinfo {volume} {8}},\ \bibinfo {pages}
  {041033} (\bibinfo {year} {2018})}\BibitemShut {NoStop}%
\bibitem [{\citenamefont {Sandvik}(2010)}]{Sandvik2010}%
  \BibitemOpen
  \bibfield  {author} {\bibinfo {author} {\bibfnamefont {A.~W.}\ \bibnamefont
  {Sandvik}},\ }in\ \href {https://doi.org/10.1063/1.3518900} {\emph {\bibinfo
  {booktitle} {{AIP} {Conference} {Proceedings}}}},\ Vol.\ \bibinfo {volume}
  {1297}\ (\bibinfo  {publisher} {AIP Publishing},\ \bibinfo {year} {2010})\
  pp.\ \bibinfo {pages} {135--338}\BibitemShut {NoStop}%
\bibitem [{\citenamefont {Sandvik}\ and\ \citenamefont
  {Evertz}(2010)}]{Sandvik10}%
  \BibitemOpen
  \bibfield  {author} {\bibinfo {author} {\bibfnamefont {A.~W.}\ \bibnamefont
  {Sandvik}}\ and\ \bibinfo {author} {\bibfnamefont {H.~G.}\ \bibnamefont
  {Evertz}},\ }\href {https://doi.org/10.1103/PhysRevB.82.024407} {\bibfield
  {journal} {\bibinfo  {journal} {Phys. Rev. B}\ }\textbf {\bibinfo {volume}
  {82}},\ \bibinfo {pages} {024407} (\bibinfo {year} {2010})}\BibitemShut
  {NoStop}%
\bibitem [{Note4()}]{Note4}%
  \BibitemOpen
  \bibinfo {note} {A different conclusion was obtained in another iPEPS
  study~\cite {xi23}, however, the results were limited to $D$ values up to 7,
  and extrapolations of $m$ were performed based on a $1/D$ extrapolation
  instead of a more accurate FCLS analysis.}\BibitemShut {Stop}%
\bibitem [{\citenamefont {Hu}\ \emph {et~al.}(2013)\citenamefont {Hu},
  \citenamefont {Becca}, \citenamefont {Parola},\ and\ \citenamefont
  {Sorella}}]{hu13}%
  \BibitemOpen
  \bibfield  {author} {\bibinfo {author} {\bibfnamefont {W.-J.}\ \bibnamefont
  {Hu}}, \bibinfo {author} {\bibfnamefont {F.}~\bibnamefont {Becca}}, \bibinfo
  {author} {\bibfnamefont {A.}~\bibnamefont {Parola}},\ and\ \bibinfo {author}
  {\bibfnamefont {S.}~\bibnamefont {Sorella}},\ }\href
  {https://doi.org/10.1103/PhysRevB.88.060402} {\bibfield  {journal} {\bibinfo
  {journal} {Phys. Rev. B}\ }\textbf {\bibinfo {volume} {88}},\ \bibinfo
  {pages} {060402} (\bibinfo {year} {2013})}\BibitemShut {NoStop}%
\bibitem [{\citenamefont {Richter}\ \emph {et~al.}(2015)\citenamefont
  {Richter}, \citenamefont {Zinke},\ and\ \citenamefont {Farnell}}]{richter15}%
  \BibitemOpen
  \bibfield  {author} {\bibinfo {author} {\bibfnamefont {J.}~\bibnamefont
  {Richter}}, \bibinfo {author} {\bibfnamefont {R.}~\bibnamefont {Zinke}},\
  and\ \bibinfo {author} {\bibfnamefont {D.~J.~J.}\ \bibnamefont {Farnell}},\
  }\href {https://doi.org/10.1140/epjb/e2014-50589-x} {\bibfield  {journal}
  {\bibinfo  {journal} {Eur. Phys. J. B}\ }\textbf {\bibinfo {volume} {88}},\
  \bibinfo {pages} {2} (\bibinfo {year} {2015})}\BibitemShut {NoStop}%
\bibitem [{\citenamefont {Wang}\ and\ \citenamefont {Sandvik}(2018)}]{wang18}%
  \BibitemOpen
  \bibfield  {author} {\bibinfo {author} {\bibfnamefont {L.}~\bibnamefont
  {Wang}}\ and\ \bibinfo {author} {\bibfnamefont {A.~W.}\ \bibnamefont
  {Sandvik}},\ }\href {https://doi.org/10.1103/PhysRevLett.121.107202}
  {\bibfield  {journal} {\bibinfo  {journal} {Phys. Rev. Lett.}\ }\textbf
  {\bibinfo {volume} {121}},\ \bibinfo {pages} {107202} (\bibinfo {year}
  {2018})}\BibitemShut {NoStop}%
\bibitem [{\citenamefont {Hering}\ \emph {et~al.}(2019)\citenamefont {Hering},
  \citenamefont {Sonnenschein}, \citenamefont {Iqbal},\ and\ \citenamefont
  {Reuther}}]{hering19}%
  \BibitemOpen
  \bibfield  {author} {\bibinfo {author} {\bibfnamefont {M.}~\bibnamefont
  {Hering}}, \bibinfo {author} {\bibfnamefont {J.}~\bibnamefont
  {Sonnenschein}}, \bibinfo {author} {\bibfnamefont {Y.}~\bibnamefont
  {Iqbal}},\ and\ \bibinfo {author} {\bibfnamefont {J.}~\bibnamefont
  {Reuther}},\ }\href {https://doi.org/10.1103/PhysRevB.99.100405} {\bibfield
  {journal} {\bibinfo  {journal} {Phys. Rev. B}\ }\textbf {\bibinfo {volume}
  {99}},\ \bibinfo {pages} {100405} (\bibinfo {year} {2019})}\BibitemShut
  {NoStop}%
\bibitem [{\citenamefont {Nomura}\ and\ \citenamefont
  {Imada}(2021)}]{nomura21}%
  \BibitemOpen
  \bibfield  {author} {\bibinfo {author} {\bibfnamefont {Y.}~\bibnamefont
  {Nomura}}\ and\ \bibinfo {author} {\bibfnamefont {M.}~\bibnamefont {Imada}},\
  }\href {https://doi.org/10.1103/PhysRevX.11.031034} {\bibfield  {journal}
  {\bibinfo  {journal} {Phys. Rev. X}\ }\textbf {\bibinfo {volume} {11}},\
  \bibinfo {pages} {031034} (\bibinfo {year} {2021})}\BibitemShut {NoStop}%
\bibitem [{\citenamefont {Ueda}\ and\ \citenamefont {Miyahara}(1999)}]{ueda99}%
  \BibitemOpen
  \bibfield  {author} {\bibinfo {author} {\bibfnamefont {K.}~\bibnamefont
  {Ueda}}\ and\ \bibinfo {author} {\bibfnamefont {S.}~\bibnamefont
  {Miyahara}},\ }\href {https://doi.org/10.1088/0953-8984/11/17/101} {\bibfield
   {journal} {\bibinfo  {journal} {J. Phys.: Condens. Matter}\ }\textbf
  {\bibinfo {volume} {11}},\ \bibinfo {pages} {L175} (\bibinfo {year}
  {1999})}\BibitemShut {NoStop}%
\bibitem [{\citenamefont {Koga}(2000)}]{koga00c}%
  \BibitemOpen
  \bibfield  {author} {\bibinfo {author} {\bibfnamefont {A.}~\bibnamefont
  {Koga}},\ }\href {https://doi.org/10.1143/JPSJ.69.3509} {\bibfield  {journal}
  {\bibinfo  {journal} {J. Phys. Soc. Jpn.}\ }\textbf {\bibinfo {volume}
  {69}},\ \bibinfo {pages} {3509} (\bibinfo {year} {2000})}\BibitemShut
  {NoStop}%
\bibitem [{\citenamefont {Vlaar}\ and\ \citenamefont
  {Corboz}(2023)}]{vlaar23b}%
  \BibitemOpen
  \bibfield  {author} {\bibinfo {author} {\bibfnamefont {P.}~\bibnamefont
  {Vlaar}}\ and\ \bibinfo {author} {\bibfnamefont {P.}~\bibnamefont {Corboz}},\
  }\href {https://doi.org/10.21468/SciPostPhys.15.4.126} {\bibfield  {journal}
  {\bibinfo  {journal} {SciPost Physics}\ }\textbf {\bibinfo {volume} {15}},\
  \bibinfo {pages} {126} (\bibinfo {year} {2023})}\BibitemShut {NoStop}%
\bibitem [{\citenamefont {Fogh}\ \emph {et~al.}(2024)\citenamefont {Fogh},
  \citenamefont {Giriat}, \citenamefont {Zayed}, \citenamefont {Piovano},
  \citenamefont {Boehm}, \citenamefont {Steffens}, \citenamefont {Safiulina},
  \citenamefont {Hansen}, \citenamefont {Klotz}, \citenamefont {Soh},
  \citenamefont {Pomjakushina}, \citenamefont {Mila}, \citenamefont {Normand},\
  and\ \citenamefont {R{\o}nnow}}]{fogh24}%
  \BibitemOpen
  \bibfield  {author} {\bibinfo {author} {\bibfnamefont {E.}~\bibnamefont
  {Fogh}}, \bibinfo {author} {\bibfnamefont {G.}~\bibnamefont {Giriat}},
  \bibinfo {author} {\bibfnamefont {M.~E.}\ \bibnamefont {Zayed}}, \bibinfo
  {author} {\bibfnamefont {A.}~\bibnamefont {Piovano}}, \bibinfo {author}
  {\bibfnamefont {M.}~\bibnamefont {Boehm}}, \bibinfo {author} {\bibfnamefont
  {P.}~\bibnamefont {Steffens}}, \bibinfo {author} {\bibfnamefont
  {I.}~\bibnamefont {Safiulina}}, \bibinfo {author} {\bibfnamefont {U.~B.}\
  \bibnamefont {Hansen}}, \bibinfo {author} {\bibfnamefont {S.}~\bibnamefont
  {Klotz}}, \bibinfo {author} {\bibfnamefont {J.-R.}\ \bibnamefont {Soh}},
  \bibinfo {author} {\bibfnamefont {E.}~\bibnamefont {Pomjakushina}}, \bibinfo
  {author} {\bibfnamefont {F.}~\bibnamefont {Mila}}, \bibinfo {author}
  {\bibfnamefont {B.}~\bibnamefont {Normand}},\ and\ \bibinfo {author}
  {\bibfnamefont {H.~M.}\ \bibnamefont {R{\o}nnow}},\ }\href
  {https://doi.org/10.1103/PhysRevLett.133.246702} {\bibfield  {journal}
  {\bibinfo  {journal} {Phys. Rev. Lett.}\ }\textbf {\bibinfo {volume} {133}},\
  \bibinfo {pages} {246702} (\bibinfo {year} {2024})}\BibitemShut {NoStop}%
\bibitem [{\citenamefont {C{\'e}pas}\ \emph {et~al.}(2001)\citenamefont
  {C{\'e}pas}, \citenamefont {Kakurai}, \citenamefont {Regnault}, \citenamefont
  {Ziman}, \citenamefont {Boucher}, \citenamefont {Aso}, \citenamefont {Nishi},
  \citenamefont {Kageyama},\ and\ \citenamefont {Ueda}}]{Cepas01}%
  \BibitemOpen
  \bibfield  {author} {\bibinfo {author} {\bibfnamefont {O.}~\bibnamefont
  {C{\'e}pas}}, \bibinfo {author} {\bibfnamefont {K.}~\bibnamefont {Kakurai}},
  \bibinfo {author} {\bibfnamefont {L.~P.}\ \bibnamefont {Regnault}}, \bibinfo
  {author} {\bibfnamefont {T.}~\bibnamefont {Ziman}}, \bibinfo {author}
  {\bibfnamefont {J.~P.}\ \bibnamefont {Boucher}}, \bibinfo {author}
  {\bibfnamefont {N.}~\bibnamefont {Aso}}, \bibinfo {author} {\bibfnamefont
  {M.}~\bibnamefont {Nishi}}, \bibinfo {author} {\bibfnamefont
  {H.}~\bibnamefont {Kageyama}},\ and\ \bibinfo {author} {\bibfnamefont
  {Y.}~\bibnamefont {Ueda}},\ }\href
  {https://doi.org/10.1103/PhysRevLett.87.167205} {\bibfield  {journal}
  {\bibinfo  {journal} {Phys. Rev. Lett.}\ }\textbf {\bibinfo {volume} {87}},\
  \bibinfo {pages} {167205} (\bibinfo {year} {2001})}\BibitemShut {NoStop}%
\bibitem [{\citenamefont {Romh{\'a}nyi}\ \emph {et~al.}(2011)\citenamefont
  {Romh{\'a}nyi}, \citenamefont {Totsuka},\ and\ \citenamefont
  {Penc}}]{romhanyi11}%
  \BibitemOpen
  \bibfield  {author} {\bibinfo {author} {\bibfnamefont {J.}~\bibnamefont
  {Romh{\'a}nyi}}, \bibinfo {author} {\bibfnamefont {K.}~\bibnamefont
  {Totsuka}},\ and\ \bibinfo {author} {\bibfnamefont {K.}~\bibnamefont
  {Penc}},\ }\href {https://doi.org/10.1103/PhysRevB.83.024413} {\bibfield
  {journal} {\bibinfo  {journal} {Phys. Rev. B}\ }\textbf {\bibinfo {volume}
  {83}},\ \bibinfo {pages} {024413} (\bibinfo {year} {2011})}\BibitemShut
  {NoStop}%
\bibitem [{\citenamefont {Boos}\ \emph {et~al.}(2019)\citenamefont {Boos},
  \citenamefont {Crone}, \citenamefont {Niesen}, \citenamefont {Corboz},
  \citenamefont {Schmidt},\ and\ \citenamefont {Mila}}]{boos19}%
  \BibitemOpen
  \bibfield  {author} {\bibinfo {author} {\bibfnamefont {C.}~\bibnamefont
  {Boos}}, \bibinfo {author} {\bibfnamefont {S.~P.~G.}\ \bibnamefont {Crone}},
  \bibinfo {author} {\bibfnamefont {I.~A.}\ \bibnamefont {Niesen}}, \bibinfo
  {author} {\bibfnamefont {P.}~\bibnamefont {Corboz}}, \bibinfo {author}
  {\bibfnamefont {K.~P.}\ \bibnamefont {Schmidt}},\ and\ \bibinfo {author}
  {\bibfnamefont {F.}~\bibnamefont {Mila}},\ }\href
  {https://doi.org/10.1103/PhysRevB.100.140413} {\bibfield  {journal} {\bibinfo
   {journal} {Phys. Rev. B}\ }\textbf {\bibinfo {volume} {100}},\ \bibinfo
  {pages} {140413} (\bibinfo {year} {2019})}\BibitemShut {NoStop}%
\bibitem [{\citenamefont {D'Emidio}\ \emph {et~al.}(2023)\citenamefont
  {D'Emidio}, \citenamefont {Eberharter},\ and\ \citenamefont
  {L{\"a}uchli}}]{demidio23}%
  \BibitemOpen
  \bibfield  {author} {\bibinfo {author} {\bibfnamefont {J.}~\bibnamefont
  {D'Emidio}}, \bibinfo {author} {\bibfnamefont {A.}~\bibnamefont
  {Eberharter}},\ and\ \bibinfo {author} {\bibfnamefont {A.}~\bibnamefont
  {L{\"a}uchli}},\ }\href {https://doi.org/10.21468/SciPostPhys.15.2.061}
  {\bibfield  {journal} {\bibinfo  {journal} {SciPost Phys.}\ }\textbf
  {\bibinfo {volume} {15}},\ \bibinfo {pages} {061} (\bibinfo {year}
  {2023})}\BibitemShut {NoStop}%
\bibitem [{\citenamefont {Czarnik}\ and\ \citenamefont
  {Corboz}(2019)}]{czarnik19b}%
  \BibitemOpen
  \bibfield  {author} {\bibinfo {author} {\bibfnamefont {P.}~\bibnamefont
  {Czarnik}}\ and\ \bibinfo {author} {\bibfnamefont {P.}~\bibnamefont
  {Corboz}},\ }\href {https://doi.org/10.1103/PhysRevB.99.245107} {\bibfield
  {journal} {\bibinfo  {journal} {Phys. Rev. B}\ }\textbf {\bibinfo {volume}
  {99}},\ \bibinfo {pages} {245107} (\bibinfo {year} {2019})}\BibitemShut
  {NoStop}%
\bibitem [{Note5()}]{Note5}%
  \BibitemOpen
  \bibinfo {note} {We do not expect this effect to persist if the VMC-PEPS data
  were extrapolated in $D$ in addition to $L$}\BibitemShut {NoStop}%
\end{thebibliography}%


\onecolumngrid

\vspace{0.5cm}

\noindent
\begin{center}
\textbf{End Matter}\\ \vspace{0.2cm}
\end{center}
\twocolumngrid

\emph{Method details.--}
Here we provide an overview of the recent methodological developments that have enabled us to perform calculations with higher precision than in previous iPEPS studies. First, optimization based on AD yields more accurate tensors than traditional imaginary-time evolution algorithms; however, it is computationally also more expensive. Recently, the efficiency of AD has been considerably improved by extending AD to the truncated singular value decomposition~\cite{francuz25} which we used to push the simulations to larger bond dimensions than in previous iPEPS studies~\cite{xi23}. Another advantage of AD is the possibility to implement the lattice symmetry of the ground state directly in the tensors, which has already been commonly done for the AF C$_{4v}$-symmetric ground state. Here we further extend this idea to represent the C$_4$ symmetric plaquette state using a $2\times2$ unit cell with a single tensor that is rotated by 90 degrees between neighboring sites in the cell. Furthermore, a key tool to obtain accurate extrapolations of order parameters in gapless phases is FCLS, which was previously successfully applied for various strongly correlated systems~\cite{corboz18,rader18,czarnik19b,hasik21,vanhecke22, hasik24}, but not to the SSM before. 

Finally, an important  technique used in our study is an accurate extrapolation of the energy as a function of the energy variance, which is computed from the contraction of a large unit cell using the corner transfer matrix renormalization group method~\cite{cortes25}, in which all relevant pairs of Hamiltonian terms contributing to the variance are evaluated in a systematic way. This recent scheme enables accurate estimates at substantially smaller contraction dimension $\chi$ than previous schemes~\cite{corboz16b,vanderstraeten16}, which is crucial for obtaining results at to large bond dimensions up to $D=10$ as  in the present study.

\emph{Extrapolation of the plaquette order parameter.--}
In Fig.~\ref{fig:pop} we present the results for the plaquette order parameter obtained with the plaquette ansatz for values of $J'/J$ around the phase transition, demonstrating that it does not vanish at the transition (see main text).

\begin{figure}[t]
  \centering
  \includegraphics[width=1\linewidth]{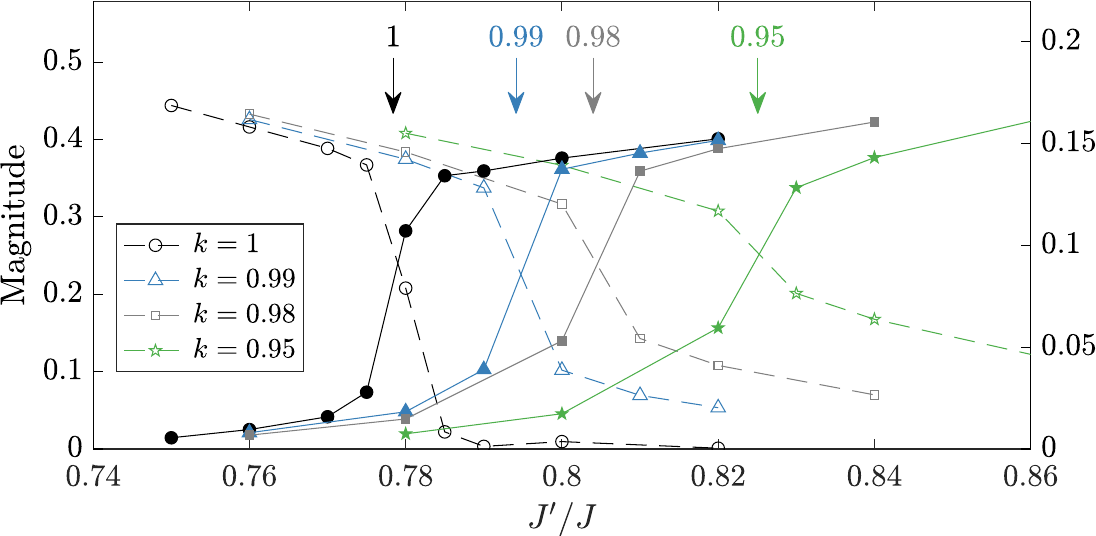}
  \caption{Plaquette order parameter $\Delta$ (open symbols, left axis) and local magnetic moment $m$ (filled symbols, right axis) as a function of $J'/J$ for a modified SSM, in which the couplings on the boundary of repeating $6\times 6 $ cells are weakened by a factor $k$, to investigate the effect of open boundary conditions.  Even a small decrease of $k$ from the isotropic limit $k=1$ shifts the phase boundary of the plaquette phase to substantially larger values of $J'/J$. The single-site setup with bond dimension $D=5$ is used.
  }
  \label{fig:OB}
\end{figure}

\emph{Sensitivity to open boundary conditions.--}
While our results for the phase boundary of the plaquette phase $J'/J= 0.785(5)$ agrees with DMRG from Ref.~\cite{yang22} ($J'/J=0.788(2)$) and Ref.~\cite{qian24} ($J'/J=0.785(5)$), ED~\cite{wang22} ($J'/J=0.789(4)$), and NQS ($J'/J\sim0.78$), it is incompatible with VMC-PEPS ($J'/J= 0.828(5)$)~\cite{liu24}.
In the latter study, the results are based on lattices with open boundary conditions at finite $D$. To gain a better understanding of why VMC-PEPS yields a different result, we study a modified SSM where we weaken the couplings by a factor $k$ at the boundary of repeating cells of size $6\times 6$, to mimic the effect of open boundary conditions. Figure~\ref{fig:OB} shows that even a small weakening of a few percent shifts the phase boundary of the plaquette phase to larger values of $J'/J \approx 0.825$, consistent with the VMC-PEPS study. 
Hence, from this qualitative analysis we infer that open boundary conditions tend to favor the plaquette phase, which, in combination with the fact that the results have not been extrapolated in $D$ in the VMC-PEPS study, can contribute to an overestimation of the extent of the plaquette phase~\footnote{We do not expect this effect to persist if the VMC-PEPS data were extrapolated in $D$ in addition to $L$}. 
This result can also be understood intuitively, since lattices of even size with open boundary conditions allow a complete covering of the lattice by plaquettes without cutting any low-energy bonds, in contrast to the AF (QSL) state. 

\begin{figure}[H]
  \centering
  \includegraphics[width=1\linewidth]{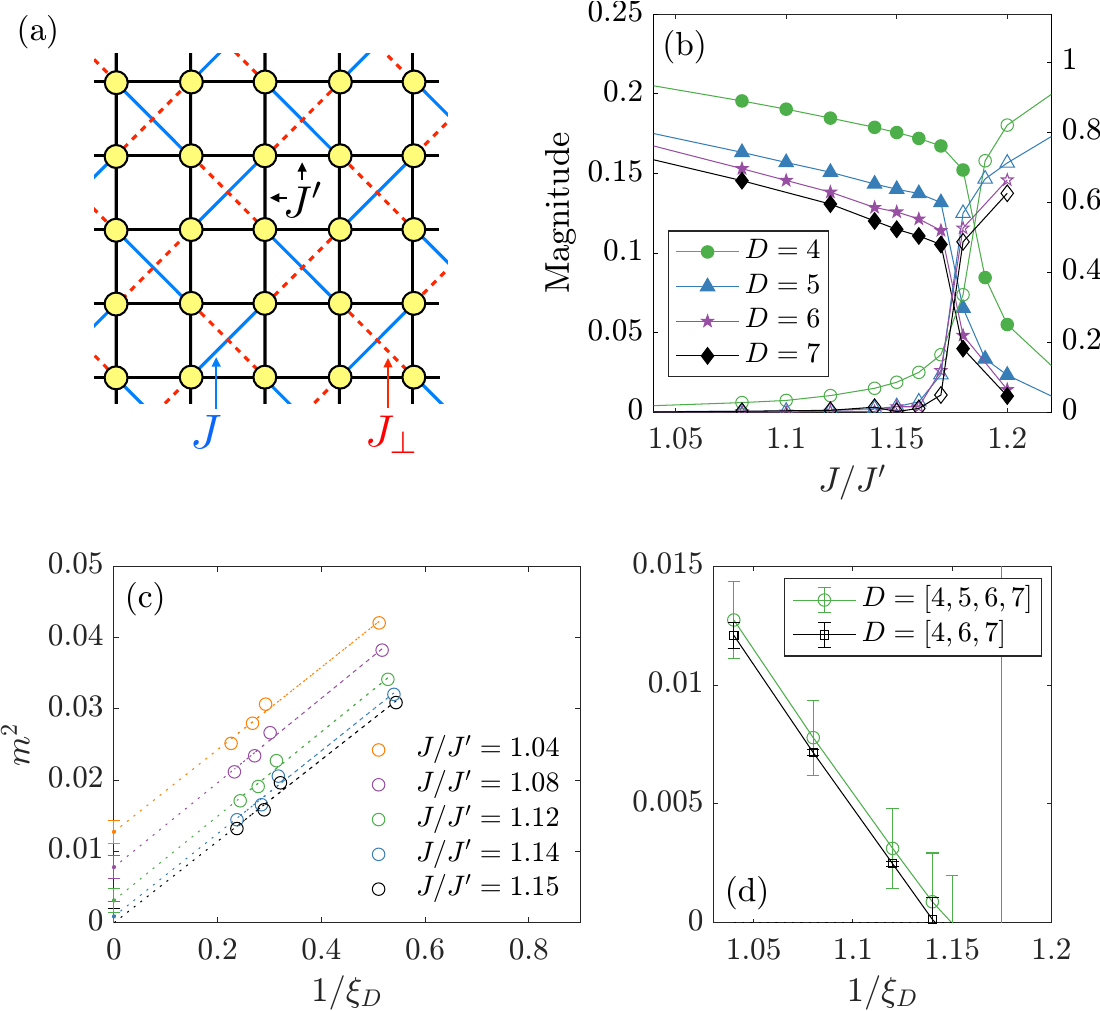}
  \caption{Results for an extended SSM with  $J_\perp/J'=0.3$, obtained with the single-site setup. (a) Description of the couplings of the extended SSM,  including an additional coupling $J_\perp$ (red dotted lines), perpendicular to the dimer coupling $J$ (blue solid lines). 
(b)   Local magnetic moment $m$ (filled symbols, left axis) and plaquette order parameter $\Delta$ (open symbols, right axis)  as a function of $J/J'$, with the phase transition located around $J/J'\sim1.175$. (c)~Linear extrapolations of the $m^2$ data as a function of the inverse effective correlation length $\xi_D$ for different values of $J/J'$. (d) Extrapolated value of $m^2$ as a function of $J/J'$, which vanishes around $J/J'\sim 1.145$, i.e. before the onset of the plaquette phase, indicated by the vertical grey line. Different sets of bond dimension values have been used for the green and black data as indicated in the legend.
  }
  \label{fig:essm}
\end{figure}

\emph{Extended SSM.--}
To check the robustness of the QSL phase we consider an extended SSM~\cite{liu24} with an additional diagonal Heisenberg interaction of strength  $J_\perp$ on the plaquettes that already have a diagonal bond, an interaction that would lead to the checkerboard lattice when $J_\perp=J$, see Fig.~\ref{fig:essm}(a). In Fig.~\ref{fig:essm}(b-d) we present data based on the single-site setup up to bond dimension $D=7$. Figure~\ref{fig:essm}(b) shows that the onset of the plaquette phase is around $J/J'\sim1.175$, whereas the FCLS analysis in Figs.~\ref{fig:essm}(c,d) reveals that the onset of the AF order is located around $J/J'\sim1.145$. 
 Hence, there is a range of values where both orders vanish, similar to the standard SSM, although the extent of the intermediate phase is smaller here, approximately half of its size at  $J_\perp=0$. These results suggest that the QSL is stable upon introducing a small $J_\perp$ coupling, at least up to a certain critical value of $J_\perp$.

\end{document}